\documentclass[a4paper,fleqn,usenatbib]{mnras}

\usepackage{newtxtext,newtxmath}

\usepackage[T1]{fontenc}
\usepackage{ae,aecompl}

\usepackage{graphicx}	
\usepackage{amsmath}	
\usepackage{amssymb}	
\usepackage{pdflscape}
\usepackage{color}

\title[Chemical Differentiation among the NGC2264 Clumps]{Investigation of Chemical Differentiation among the NGC2264 Cluster-Forming Clumps}

\author[K. Taniguchi et al.]{
Kotomi Taniguchi,$^{1,2,3}$\thanks{E-mail: kotomi.taniguchi@gakushuin.ac.jp}
Adele Plunkett,$^{4}$
Eric Herbst,$^{1,2}$
Kazuhito Dobashi,$^{5}$
\newauthor{
Tomomi Shimoikura,$^{6}$
Fumitaka Nakamura,$^{7,8,9}$
Masao Saito$^{7,8}$}
\\
$^{1}$Department of Astronomy, University of Virginia, Charlottesville, VA 22904, USA\\
$^{2}$Department of Chemistry, University of Virginia, Charlottesville, VA 22903, USA\\
$^{3}$Department of Physics, Faculty of Science, Gakushuin University, 1-5-1, Mejiro, Toshima, Tokyo 171-8588, Japan\\
$^{4}$National Radio Astronomy Observatory, 520 Edgemont Rd., Charlottesville, VA 22903, USA \\
$^{5}$Department of Astronomy and Earth Sciences, Tokyo Gakugei University, Nukuikitamachi, Koganei, Tokyo 184-8501, Japan\\
$^{6}$Faculty of Social Information Studies, Otsuma Women's University, Sanban-cho, Chiyoda, Tokyo 102-8357, Japan\\
$^{7}$National Astronomical Observatory of Japan (NAOJ), Osawa, Mitaka, Tokyo 181-8588, Japan\\
$^{8}$Department of Astronomical Science, School of Physical Science, SOKENDAI (The Graduate University for Advanced Studies), Osawa, Mitaka,\\Tokyo 181-8588, Japan\\
$^{9}$The University of Tokyo, Hongo, Bunkyo, Tokyo 113-0033, Japan\\
}

\pubyear{2019}

\begin{document}
\label{firstpage}
\pagerange{\pageref{firstpage}--\pageref{lastpage}}
\maketitle

\begin{abstract}
We have carried out mapping observations of molecular emission lines of HC$_{3}$N and CH$_{3}$OH toward two massive cluster-forming clumps, NGC2264-C and NGC2264-D, using the Nobeyama 45-m radio telescope.
We derive an $I$(HC$_{3}$N)/$I$(CH$_{3}$OH) integrated intensity ratio map, showing a higher value at clumps including 2MASS point sources at the northern part of NGC2264-D.
Possible interpretations of the $I$(HC$_{3}$N)/$I$(CH$_{3}$OH) ratio are discussed.
We have also observed molecular emission lines from CCS and N$_{2}$H$^{+}$ toward five positions in each clump.
We investigate the $N$(N$_{2}$H$^{+}$)/$N$(CCS) and $N$(N$_{2}$H$^{+}$)/$N$(HC$_{3}$N) column density ratios among the ten positions in order to test whether they can be used as chemical evolutionary indicators in these clumps.
The $N$(N$_{2}$H$^{+}$)/$N$(CCS) ratio shows a very high value toward a bright embedded IR source (IRS1), whereas the $N$(N$_{2}$H$^{+}$)/$N$(HC$_{3}$N) ratio at IRS1 is comparable with those at the other positions.
These results suggest that UV radiation affects the chemistry around IRS1.
We find that there are positive correlations between these column density ratios and the excitation temperatures of N$_{2}$H$^{+}$, which implies the chemical evolution of clumps.
These chemical evolutionary indicators likely reflect the combination of evolution along the filamentary structure and evolution of each clump.
\end{abstract}




\section{Introduction} \label{sec:intro}

The majority of stars are born in clusters \citep[e.g.,][]{2003ARA&A..41...57L} and recent studies showed evidence that our Sun was also born in such a cluster region \citep{2010ARA&A..48...47A}.
Hence, it is essential for our understanding of environments where star formation occurs and the solar system was born to study cluster regions.
Despite the importance of cluster regions, formation and evolutionary processes of clusters are still poorly known.

Young clusters are identified in near-infrared images, and they are often embedded in their natal molecular clumps \citep[e.g.,][]{2011PASJ...63S...1D, 2013ApJ...768...72S}.
Therefore, not only infrared observations but also millimeter and submillimeter observations, which trace molecular rotational transition lines and dust continuum emission, are important to understand cluster-forming regions. 
Recently, \citet{2018ApJ...855...45S} conducted observations toward 15 regions with the Nobeyama 45-m radio telescope and identified 24 clumps in massive cluster-forming regions.
Based on the spatial coincidence between the gas and the star density, they categorized  cluster-forming clumps into four types: Clumps without clusters (Type 1), clumps showing good correlations with clusters (Type 2), clumps showing poor correlations with clusters (Type 3), and clusters with no associated clumps (Type 4). 
They proposed that the cluster formation proceeds from Type 1 through Types 2 and 3, to Type 4.

Nowadays,  chemical composition is known as a useful tool to investigate the evolution and physical conditions during star formation \citep[e.g.,][]{2012A&ARv..20...56C,2018IAUS..332....3V}.
For example, some chemical evolutionary indicators have been proposed for low-mass star-forming cores, combining  N-bearing species (NH$_{3}$ and N$_{2}$H$^{+}$) with carbon-chain molecules, which are known as late-type and early-type species, respectively \citep{1992ApJ...392..551S,1998ApJ...506..743B}.
In high-mass star-forming regions, the $N$(N$_{2}$H$^{+}$)/$N$(HC$_{3}$N) column density ratio was found to be a chemical evolutionary indicator based on the results of the observations with the Nobeyama 45-m radio telescope \citep{2019ApJ...872..154T}.

As another advantage to chemical composition as a tool, the chemical differentiation around protostars seems to reflect the variety of star formation processes \citep{2013ChRv..113.8981S,2019ApJ...881..57T} and environments \citep{2016A&A...592L..11S}.
Saturated complex organic molecules (COMs), consisting of more than 6 atoms, have been found to be abundant around massive young protostars, produced by hot core chemistry \citep[e.g.,][]{2009ARA&A..47..427H}.
Similar chemical compositions have been discovered around low-mass protostars, caused by what is known as hot corino chemistry.
COMs are formed on dust grains during the cold-prestellar and warm-up phases \citep{2006A&A...457..927G}, and efficiently sublimate into the gas phase during heating from protostars. 
Further gas-phase reactions also produce COMs \citep[e.g.,][]{2019MNRAS.482.3567S}.

In contrast to COMs, carbon-chain molecules are classically known as early-type species, because they are abundant in young starless cores and deficient in star-forming cores \citep{1992ApJ...392..551S}.
They are formed from atomic carbon (C) and ionic carbon (C$^{+}$) in the gas phase via ion-molecule and neutral-neutral reactions.
However, recent studies have shown that carbon-chain molecules can be regenerated around protostars \citep{2008ApJ...681.1385H}, where the reaction between CH$_{4}$ sublimated from dust grains and C$^{+}$ initiates carbon-chain formation.
Such a carbon-chain chemistry was first found around low-mass protostars and named Warm Carbon Chain Chemistry \citep[WCCC;][]{2013ChRv..113.8981S}.
Recent observations toward massive young stellar objects (MYSOs) also showed that cyanopolyynes (HC$_{2n+1}$N) are abundant around some MYSOs \citep{2017ApJ...844...68T,2018ApJ...866..150T}.
In addition, HC$_{3}$N appears to be formed in the warm dense gas around high-mass protostellar objects \citep{2018ApJ...854..133T,2019ApJ...872..154T}. 
\citet{2018ApJ...866..150T} found that COM-poor cores are surrounded by cyanopolyyne-rich envelopes and COM-rich cores, namely hot cores, are surrounded by CH$_{3}$OH-rich envelopes.

The chemical differentiation between carbon-chain-rich conditions and COM-rich conditions may be caused by the different timescales  for the duration of the prestellar collapse stage \citep{2013ChRv..113.8981S}.
In particular,  long and short prestellar collapse stages lead to COM-rich and carbon-chain-rich conditions, respectively.  
If there is enough time for ionic/atomic carbon to convert into CO molecules, CO molecules are adsorbed on to dust grains and the hydrogenation of CO on dust surface generates CH$_{3}$OH and COMs abundantly.
On the other hand, carbon atoms are depleted on to dust grains during the short timescale of prestellar collapse, and CH$_{4}$ can be efficiently formed on dust surfaces. 

Another possible cause of the origin of the chemical differentiation around protostars is the penetration of the interstellar radiation field \citep{2016A&A...592L..11S}.
The interstellar radiation field destroys CO molecules to form atomic carbon and oxygen.
Therefore, CH$_{3}$OH is rich in the well shielded regions, while carbon-chain molecules are abundant in the regions irradiated by the interstellar radiation field. Recently, different timescales of the warm-up phase were proposed as another possible origin of the chemical differentiation \citep{2019ApJ...881..57T}.  
The fast and slow warm-up phases will produce CH$_{3}$OH-rich and cyanopolyyne-rich conditions, respectively.
This timescale should depend on the size of warm region and the infall velocity, which are related to various physical conditions in star-forming regions.

Although the examples given above are for individual stars on core-scale in low-mass star-forming regions and on clump-scale in high-mass star-forming regions, such relationships have been confirmed by observations.
In this paper, we investigate the chemical composition in the NGC2264-C and NGC2264-D cluster-forming clumps  and test the utility of chemical composition as a probe on larger scales than studied before.  
NGC2264 is located in the Mon OB1 association, the distance of which is approximately 760 pc \citep{2008hsf1.book..966D}.
Both of the cluster-forming clumps are classified as Type 2 by \citet{2018ApJ...855...45S}, which means that clusters correlate with the clumps.
Regarding structure in these clumps,  there are two well-characterized IR sources surrounded by a variety of young stellar objects (YSOs).
IRAS 06384+0932, also referred to as IRS1,  is a bright embedded IR source in the NGC2264-C clump.
This source is a B2-type object with bolometric luminosity of $L_{\rm {bol}} \sim 2300$ L$_{\sun}$ \citep{2006A&A...445..979P}.
IRAS 06382+0939, which is also called IRS2, is a Class I YSO with $L_{\rm {bol}} \sim 150$ L$_{\sun}$ in the NGC2264-D clump \citep{2006A&A...445..979P}.

In Section \ref{sec:obs}, we describe our observations.
The resultant integrated intensity maps and spectra are shown in Sections \ref{sec:res1} and \ref{sec:res2}, respectively, while the analyses and results are summarised in Section \ref{sec:ana}.
As mentioned in the above, the [N-bearing species]/[carbon-chain species] ratios are considered to be chemical evolutionary indicators, and the [carbon-chain species]/[COMs] ratios seem to reflect a variety of the environments or formation processes of cores/clumps.
We discuss our investigation of the $N$(N$_{2}$H$^{+}$)/$N$(CCS) and $N$(N$_{2}$H$^{+}$)/$N$(HC$_{3}$N) column density ratios in Section \ref{sec:d1} and the $I$(HC$_{3}$N)/$I$(CH$_{3}$OH) integrated intensity ratio in Section \ref{sec:d2}.
The conclusions of this paper are summarised in Section \ref{sec:con}.

\section{Observations and data reduction} \label{sec:obs}

All the observations described in this paper were carried out in 2019 January with the Nobeyama 45-m radio telescope (Proposal ID: CG181002, PI: Kotomi Taniguchi, 2018-2019 season).
Table~\ref{tab:line1} summarises the rest frequency and excitation energy of the observed lines \citep[CDMS;][]{2005JMoSt.742..215M}. 
We could not detect CCS ($J_{N}=7_{7}-6_{6}$) in the mapping observations with the current sensitivity, and we will not discuss it in the following sections.
The HNC ($J=1-0$) line was simultaneously observed in the same spectrometer as the CCS line ($J_{N}=7_{7}-6_{6}$).
Other CH$_{3}$OH emission lines, $2_{0}-1_{0}$, A$^{+}$ and $2_{-1}-1_{-1}$, E, were observed in the same frequency band of the $2_{0}-1_{0}$, E transition line of CH$_{3}$OH.
We categorize these three CH$_{3}$OH lines as the low-excitation-energy lines.
On the other hand, the $8_{0}-7_{1}$, A$^{+}$ transition of CH$_{3}$OH has a high excitation energy, and we call this line the high-excitation-energy line in the following sections.

\begin{table}
	\centering
	\caption{Summary of the observed lines}
	\label{tab:line1}
	\begin{tabular}{llcc} 
		\hline
		Species & Transition & Rest Freq. & $E_{\rm u}/k$ \\
		            &                 & (GHz)        & (K) \\
		\hline
		\multicolumn{4}{l}{Mapping Observations} \\
		HC$_{3}$N & $J=10-9$ & 90.979023 & 24.0 \\
		CH$_{3}$OH & $8_{0}-7_{1}$, A$^{+}$ & 95.169391 & 83.5$^{a}$ \\
		CH$_{3}$OH & $2_{0}-1_{0}$, E & 96.744545 & 20.1$^{b}$ \\
		CH$_{3}$OH & $2_{0}-1_{0}$, A$^{+}$ & 96.741371 & 6.97$^{b}$ \\
		CH$_{3}$OH & $2_{-1}-1_{-1}$, E & 96.739358 & 12.5$^{b}$ \\
		CCS & $J_{N}=7_{7}-6_{6}$ & 90.686381 & 26.1 \\
		HNC & $J=1-0$ & 90.663568 & 4.4 \\
		\multicolumn{4}{l}{Position-switching Observations} \\
		N$_{2}$H$^{+}$ & $J=1-0$ & 93.1733977 & 4.47  \\
		CCS & $J_{N}=8_{7}-7_{6}$ & 93.8701070 & 19.9  \\
		\hline
		\multicolumn{4}{l}{Note: $a$ and $b$ refer to our characterization of these}\\
		\multicolumn{4}{l}{lines into high-excitation-energy line low-excitation-} \\
		\multicolumn{4}{l}{energy, respectively.}
	\end{tabular}
\end{table}

\subsection{Mapping Observations} \label{sec:obs1}

The mapping observations of HC$_{3}$N ($J=10-9$) and CH$_{3}$OH ($8_{0}-7_{1}$, A$^{+}$ and $2_{0}-1_{0}$, E) lines toward NGC2264-C (hereafter the South region) and -D (hereafter the North region) were carried out using the 4-beam, 2-polarization, 2-sideband FOREST receiver \citep{2016SPIE.9914E..1ZM}.
The main beam efficiency ($\eta_{\rm {MB}}$) and beam size (HPBW) at 96 GHz were 51\% and 17.3\arcsec, respectively.
The system temperatures were 140 -- 170 K during the observations.

We used the SAM45 FX-type digital correlator in frequency setting with bandwidth and frequency resolution of 62.5 MHz and 30.52 kHz, respectively, applying the spectral window mode.
The frequency resolution corresponds to a velocity resolution of 0.1\,km\,s$^{-1}$.

We employed the On-The-Fly (OTF) observing mode.
The mapping areas were $6\arcmin \times 6\arcmin$ for both regions and the grid spacing was set to 7.5\arcsec.
The center coordinates of the maps were ($\alpha_{2000}$, $\delta_{2000}$) = ($6^{\rm {h}}41^{\rm {m}}$11\fs124, +9\degr29\arcmin00\farcs28) for the South region and ($6^{\rm {h}}41^{\rm {m}}$04\fs157, +9\degr34\arcmin29\farcs06) for the North region.
The off-source positions were determined based on the CO maps \citep{2018ApJ...855...45S}.
The coordinates of the off-source positions were ($\alpha_{2000}$, $\delta_{2000}$) = ($6^{\rm {h}}40^{\rm {m}}55^{\rm {s}}$, +9\degr29\arcmin00\arcsec) for the South region and ($6^{\rm {h}}41^{\rm {m}}25^{\rm {s}}$, +9\degr35\arcmin00\arcsec) for the North region.

The telescope pointing was checked every 1.5 -- 2 hr by observing the SiO ($J=1-0$) maser line from  Orion-KL at ($\alpha_{2000}$, $\delta_{2000}$) = ($5^{\rm {h}}35^{\rm {m}}$14\fs5, -5\degr22\arcmin30\farcs4).
We used the H40 receiver for the pointing observations, and found that the pointing error was within $3\arcsec$.

The data reduction was conducted with the NOSTAR, which is the software for data reduction of the OTF data obtained with the Nobeyama 45-m telescope \citep{2008PASJ...60..445S}.
The effective angular resolution of the resultant maps is $21\arcsec$.

\subsection{Position-switching Observations} \label{sec:obs2}

Based on the integrated intensity map of HC$_{3}$N obtained by the mapping observations (Section \ref{sec:obs1}), we chose 5 positions in each region, which show strong HC$_{3}$N emission (see Figs.~\ref{fig:mom0}c and \ref{fig:mom0}e).
Table~\ref{tab:position} summarises the coordinates of each position, plus the number of associated YSOs identified by 2MASS or {\it Spitzer} data \citep{2003yCat.2246....0C, 2014ApJ...794..124R} and their names in the Simbad identifier\footnote{\url{http://simbad.u-strasbg.fr/simbad/}}.
The s1 -- s3 positions are located in the strong HC$_{3}$N emission peak around IRS1, while the s4 and s5 positions are at eastern positions in the elongated structure.
The n1 position is the eastern edge of the filamentary structure, while the n2 -- n4 positions are located at the north-west parts showing a ring-like structure, and the n5 position is an isolated strong HC$_{3}$N peak in the south-east part in the North region.
The off-source positions are the same as the OTF observations (Section \ref{sec:obs1}).

The position-switching observations were carried out using the FOREST receiver \citep{2016SPIE.9914E..1ZM} with the single-beam mode.
The main beam efficiency and beam size (HPBW) at 93 GHz were 53\% and 17.8\arcsec, respectively.
The system temperatures were 120 -- 200 K.
The SAM45 correlator is run with bandwidth and frequency resolution of 125 MHz and 30.52 kHz, respectively, without the spectral window mode, and the frequency resolution corresponds to the velocity resolution of 0.1\,km\,s$^{-1}$.
We conducted the 2-channel binning in the final spectra, and then the velocity resolution of the final spectra is smoothed to 0.2\,km\,s$^{-1}$.

We set the scan pattern at 20 s for both on-source and off-source positions.
The chopper-wheel calibration method was employed and the absolute calibration error was around 10\%.
The procedure of the pointing observations were the same as the mapping observations (Section \ref{sec:obs1}), and the pointing error was less than $3\arcsec$.

We conducted data reduction using the Java Newstar, which is the astronomical data analyzing system of the single-position observations obtained with the Nobeyama 45-m telescope.
The total on-source integration times are between 20 min and 45 min.

\begin{landscape}
\begin{table}
	\centering
	\caption{Positions of the position-switching observations and associated YSOs}
	\label{tab:position}
	\begin{tabular}{ccccl} 
		\hline
		Position name & R.A. (J2000) & Decl. (J2000) & Num. of YSOs$^{a}$ & Name of YSOs$^{b}$ \\
		\hline
		\multicolumn{5}{l}{NGC2264-C (South)} \\
		s1 & $06^{\rm {h}}41^{\rm {m}}$10\fs99 & +9\degr29\arcmin21\farcs5 & 0$^{c}$ & ... \\
		s2 & $06^{\rm {h}}41^{\rm {m}}$09\fs16 & +9\degr29\arcmin38\farcs7 & 3 & J06410890+0929451, Allen's Source (IRS1), J06410954+0929250 \\
		s3 & $06^{\rm {h}}41^{\rm {m}}$12\fs87 & +9\degr29\arcmin08\farcs1 & 3 & J06411259+0929048, J06411196+0929089, [RPG2014] 28064 \\
		s4 & $06^{\rm {h}}41^{\rm {m}}$14\fs51 & +9\degr28\arcmin34\farcs0 & 2 & [RPG2014] 27511, [RPG2014]27392 \\
		s5 & $06^{\rm {h}}41^{\rm {m}}$15\fs24 & +9\degr29\arcmin07\farcs2 & 1 & [RPG2014] 27970 \\
		\multicolumn{5}{l}{NGC2264-D (North)} \\
		n1 & $06^{\rm {h}}41^{\rm {m}}$12\fs42 & +9\degr35\arcmin29\farcs4 & 2 & J06411184+0935313, [RPG2014] 32898 \\
		n2 & $06^{\rm {h}}41^{\rm {m}}$01\fs03 & +9\degr35\arcmin47\farcs3 & 1 & J06410028+0935591 \\
		n3 & $06^{\rm {h}}41^{\rm {m}}$00\fs16 & +9\degr36\arcmin18\farcs0 & 4 & J06410063+0936103, J06405952+0936105, J06410025+0936311, J06405981+0936332 \\
		n4 & $06^{\rm {h}}40^{\rm {m}}$57\fs74 & +9\degr35\arcmin33\farcs4 & 1 & [RPG2014] 33125\\
		n5 & $06^{\rm {h}}41^{\rm {m}}$06\fs52 & +9\degr34\arcmin03\farcs1 & 3 & J06410665+0933576, J06410629+0933496, J06410562+0933549 \\
		\hline
		\multicolumn{5}{l}{$a$ Number of YSOs within a $18 \arcsec$ beam size, corresponding to the beam size at 93 GHz.} \\
		\multicolumn{5}{l}{$b$ The names were taken from the Simbad identifier. Objects with brackets of [PRG2014] mean that they were identified by {\it Spitzer} data \citep{2014ApJ...794..124R}.}\\
		\multicolumn{5}{l}{Others starting with "J" were identified by 2MASS data \citep{2003yCat.2246....0C}.} \\
		\multicolumn{5}{l}{$c$ Non-detection of any YSOs at the s1 position may be caused by the nearby bright IRS1 source.}
	\end{tabular}
\end{table}
\end{landscape}

\section{Results and Analyses} \label{sec:resana}

\subsection{Results of Mapping Observations} \label{sec:res1}

\begin{figure*}
       \centering
	\includegraphics[bb=10 50 449 800, scale = 0.82]{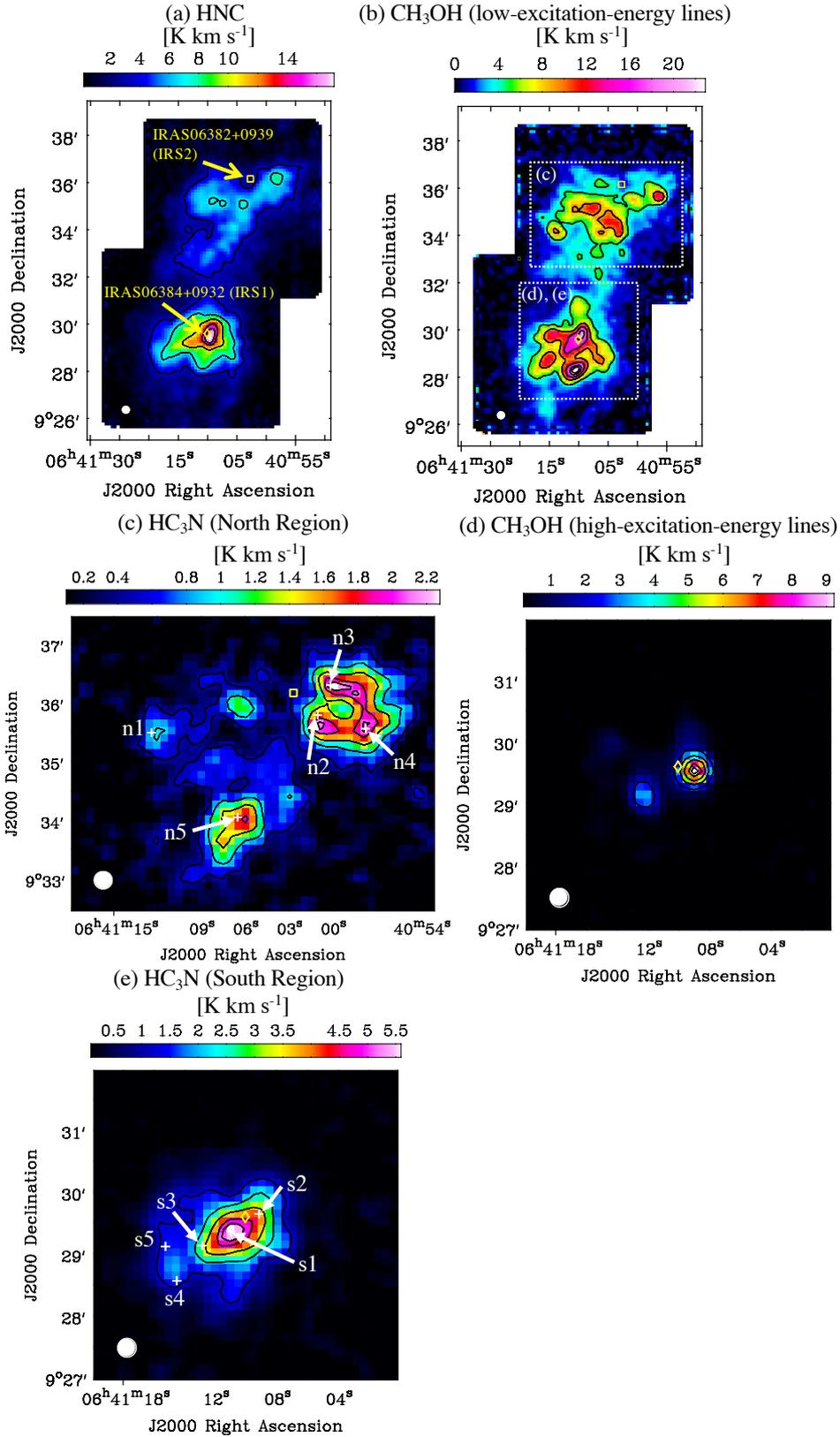}
       \caption{Integrated Intensity maps of (a) HNC ($J=1-0$), (b) CH$_{3}$OH (low-excitation-energy lines), (c) HC$_{3}$N in the North region, (d) CH$_{3}$OH ($8_{0}-7_{1}$ A$^{+}$) in the South region, and (e) HC$_{3}$N in the South region.
    The contour levels are 0.2, 0.4, 0.6, 0.8, and 0.9 of their peak levels, where the peak intensities are 17.2,  23.0, 2.5, 8.9, and 5.5 K \,km\,s$^{-1}$ for (a), (b), (c), (d), and (e), respectively.
    The yellow diamond and square indicate the IRAS06384+0932 (IRS1) and IRAS06382+0939 (IRS2) positions, respectively.
    The white dashed squares in panel (b) indicate the regions of panels (c), (d) and (e).
    Panels (c) and (e) show the positions of the position-switching observations (n1 -- n5 and s1 -- s5) as indicated by the white cross marks. 
    The white filled circles indicate the beam size of the Nobeyama 45-m telescope at 93 GHz ($\approx 18$\arcsec).}
    \label{fig:mom0}
\end{figure*}

Table~\ref{tab:mom0} summarises the noise levels of the integrated intensity maps which are shown in Fig.~\ref{fig:mom0}.
The HNC line is a good dense gas tracer, and we show its integrated intensity map in Fig.~\ref{fig:mom0}a.
The spatial distribution of HNC seems to be filamentary and more fragmented in the North region, which resembles that of another dense gas tracer N$_{2}$H$^{+}$ ($J=1-0$) \citep[see Fig.~5 of][]{2006A&A...445..979P}.
\citet{2002ApJ...568..259W} suggested that such a fragmented structure with low-intensity peak implies a slightly more evolved cluster in which individual protostars start to shed their circumstellar envelopes and migrate away from their birthplace, based on the distribution of the 870 $\mu$m dust continuum emission.
On the other hand, in the South region, the HNC distribution shows a core-like structure centred nearly at IRS1.
The morphology is similar to the N$_{2}$H$^{+}$ and the 1.2mm dust continuum emission \citep[see Fig.~4 of][]{2006A&A...445..979P}.

\begin{table}
	\centering
	\caption{Qualities of integrated intensity maps}
	\label{tab:mom0}
	\begin{tabular}{cccc} 
		\hline
		Panel & Species & Transition & noise level \\
		          &   &                 & (K \,km\,s$^{-1}$) \\
		\hline
		(a) & HNC & $J=1-0$ & 0.2 \\
		(b) & CH$_{3}$OH & $2_{0}-1_{0}$, E & 0.4 \\
		(c), (e) & HC$_{3}$N & $J=10-9$ & 0.1 \\
		(d) & CH$_{3}$OH & $8_{0}-7_{1}$, A$^{+}$ & 0.1 \\
		\hline
	\end{tabular}
\end{table}

The spatial distribution of the low-excitation-energy lines of CH$_{3}$OH is shown in Fig.~\ref{fig:mom0}b.
As shown in Fig.~\ref{fig:CH3OHlow} in Appendix \ref{sec:appen}, there are three lines in the narrow frequency range.
At some positions, especially near IRS1, the line widths are so wide that we cannot make integrated intensity maps of each line. 
Hence, Fig.~\ref{fig:mom0}b shows the integrated intensity map of the sum of these three lines.
The spatial distribution of the low-excitation-energy CH$_{3}$OH line shows more complicated structures compared to that of HNC.
Its distribution is similar to that of SO ($J_{N}=3_{2}-2_{1}$) \citep[see Fig.~6 of][]{2018ApJ...855...45S}, which may suggest that the low-excitation-energy CH$_{3}$OH originates partially from shock regions rather than hot core regions \citep[e.g.,][]{2007prpl.conf..245A}.
On the other hand, the high-excitation-energy line of CH$_{3}$OH is detected only near IRS1 (Fig.~\ref{fig:mom0}d and Fig.~\ref{fig:CH3OHhigh}).
This line is considered to originate from the hot gas around protostars, normally referred to as hot cores \citep{2015ApJ...809..162W,2017ApJ...847..108W}.

Fig.~\ref{fig:mom0}c shows the spatial distribution of HC$_{3}$N in the North region.
The ring-like structure with three prominent peaks can be seen at the north-west part, and are labelled n2 -- n4.
Another prominent peak corresponds to the n5 position.
This position was identified as D-MM1 by \citet{2006A&A...445..979P}, who mentioned that this position contains a Class 0 object.
In the South region (Fig.~\ref{fig:mom0}e), the HC$_{3}$N emission is concentrated around IRS1.
Furthermore, the HC$_{3}$N emission is stronger and its line widths are wider around IRS1 (Fig.~\ref{fig:HC3N}). 
HC$_{3}$N is likely to form and/or to survive in the warm dense gas around the protostars \citep{2008ApJ...681.1385H, 2019ApJ...881..57T}.

\subsection{Results of Position-Switching Observations} \label{sec:res2}

\begin{figure*}
	\includegraphics[width=12cm]{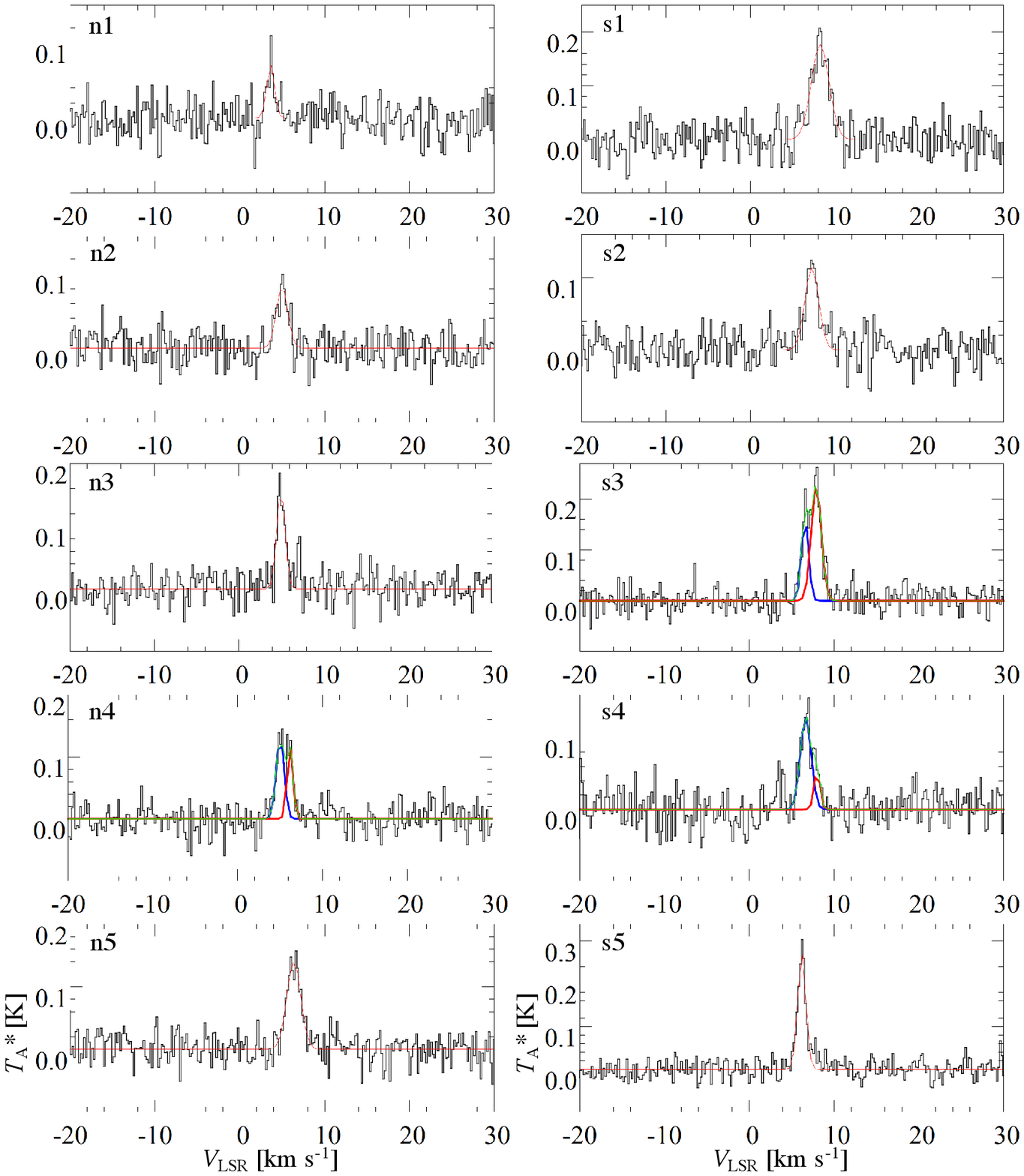}
    \caption{Spectra of CCS ($J_{N}=8_{7}-7_{6}$) toward n1 -- n5 and s1 -- s5 obtained with the position-switching observations with the Nobeyama 45-m telescope. 
    The red curves indicate the results of the Gaussian fitting. 
    In the case of n4, s3, s4, we applied two-velocity-component Gaussian fitting indicated by the blue and red curves, and the synthesized spectra of the two components are indicated by the green curves.}
    \label{fig:CCS}
\end{figure*}

Fig.~\ref{fig:CCS} shows spectra of CCS ($J_{N}=8_{7}-7_{6}$) toward the 10 positions.
The rms noise levels are 19 -- 22 mK for all of the positions except for s1 where the rms noise level is 26 mK.
The CCS line has been detected with a signal-to-noise (S/N) ratio above 4 toward all of the positions.
We fitted the spectra with a Gaussian profile and obtained the spectral line parameters shown in Table~\ref{tab:gaussCCS}.
In the case of the n4, s3, and s4 positions, we applied two-velocity-component Gaussian fitting.
In Fig.~\ref{fig:CCS}, the blue and red curves indicate each component, and their synthesized profiles are shown by the green curves.

We also produced spectra of the HC$_{3}$N line from the OTF data, as shown in Fig.~\ref{fig:HC3N}. 
The spectral line parameters for these lines are summarised in Table~\ref{tab:gaussCCS}.

\begin{figure*}
	\includegraphics[width=12cm]{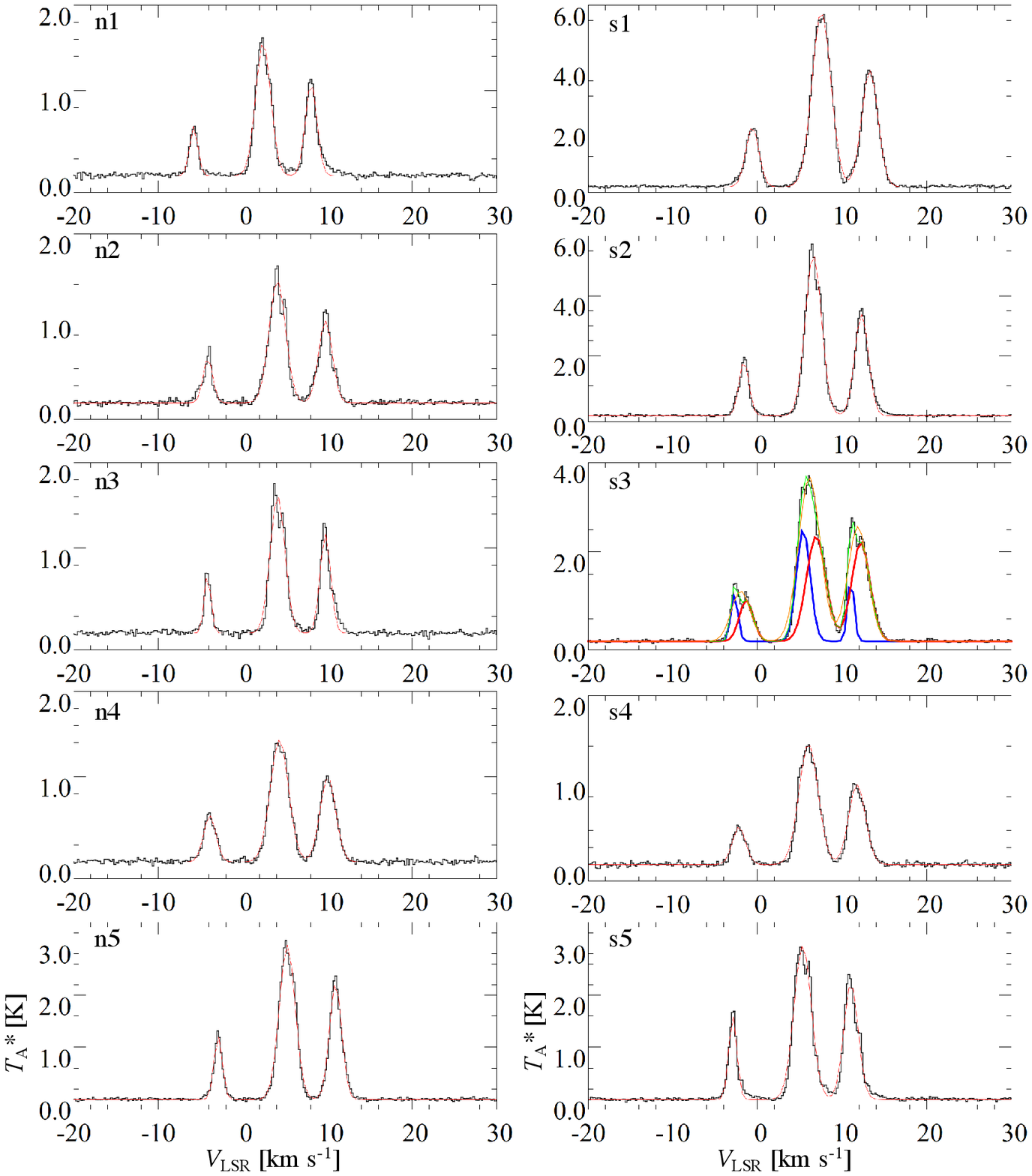}
    \caption{Spectra of N$_{2}$H$^{+}$ ($J=1-0$) toward n1 -- n5 and s1 -- s5 obtained with the position-switching observations with the Nobeyama 45-m telescope. 
    The red curves indicate the results of the Gaussian fitting. 
    In the case of s3, we applied two-velocity-component Gaussian fitting indicated by the blue and red curves, and the synthesized spectra of the two components are indicated by the green curves. The orange curve shows the fitting result of the single-velocity-component Gaussian fitting.}
    \label{fig:N2H}
\end{figure*}

Fig.~\ref{fig:N2H} shows spectra of N$_{2}$H$^{+}$ ($J=1-0$) at the 10 positions.
The emission has been detected with an S/N ratio above 4 toward all the positions.
The rms noise levels are 19 -- 22 mK for all of the positions except for s1 where the rms noise level is 27 mK.
We performed a Gaussian fitting and obtained the spectral line parameters as summarised in Table~\ref{tab:gaussN2H}. 
The two-velocity-component Gaussian fitting was applied to s3, and the blue and red curves indicate each component, and the green curve is the synthesized profile.
The orange curve indicates the result of one-velocity-component Gaussian fitting.
We used the value obtained with this fitting when we derived its column density and excitation temperature (Section~\ref{sec:ana}).

Several spectra exhibit features of molecular outflows \citep{2012MNRAS.423.1127B}. 
For example, outflow wings are seen in the spectra of HC$_{3}$N at the positions s2, s5, and n2 in Fig.~\ref{fig:HC3N}.
Multiple velocity components are seen in CCS at s3, s4, and n4 in Fig.~\ref{fig:CCS}, and in N$_{2}$H$^{+}$ at s3 in Fig.~\ref{fig:N2H}.

\begin{table}
	\centering
	\caption{Spectral line parameters of CCS ($J_{N}=8_{7}-7_{6}$) and HC$_{3}$N ($J=10-9$) at 10 positions}
	\label{tab:gaussCCS}
	\begin{tabular}{cccccc} 
		\hline
		Position  & $V_{\rm {LSR}}$ & $T_{\rm {mb}}$  & $\Delta v$ & $\int T_{\rm {mb}} dv$ & rms \\
		    &  (\,km\,s$^{-1}$) &  (K) & (\,km\,s$^{-1}$)   & (K \,km\,s$^{-1}$) & (mK) \\     
		\hline
		\multicolumn{6}{c}{\textbf{CCS ($J_{N}=8_{7}-7_{6}$)}} \\
		n1 & 3.64 (9) & 0.14 (2) & 1.1 (2) & 0.16 (4) & 19.7 \\
		n2 & 4.99 (9) & 0.16 (2) & 1.6 (2) & 0.27 (5) & 19.1 \\
		n3 & 4.98 (5) & 0.31 (3) & 1.15 (12) & 0.38 (5) & 20.7 \\
		n4 & 5.01 (9) & 0.26 (2) & 1.1 (2) & 0.30 (6) & 21.2 \\
		     & 6.18 (8) & 0.22 (3) & 0.81 (17) & 0.19 (5) & ... \\
		n5 & 6.37 (6) & 0.27 (2) & 1.95 (15) & 0.56 (6) & 21.7 \\
		s1 & 8.27 (7) & 0.34 (2) & 2.47 (18) & 0.90 (8) & 25.8 \\
		s2 & 7.27 (8) & 0.22 (2) & 1.94 (19) & 0.45 (6) & 20.5 \\
		s3 & 6.57 (8) & 0.31 (4) & 0.98 (15) & 0.33 (6) & 21.2 \\
		     & 7.91 (8) & 0.44 (2) & 1.39 (16) & 0.65 (8) & ... \\
		s4 & 6.67 (11) & 0.25 (2) & 1.5 (3) & 0.40 (8) & 21.7 \\
		     & 8.00 (18) & 0.10 (4) & 0.8 (4) & 0.09 (6) &  ... \\
		s5 & 6.20 (2) & 0.52 (2) & 1.14 (6) & 0.63 (4) & 21.0 \\ 
		\multicolumn{6}{l}{} \\
		\multicolumn{6}{c}{\textbf{HC$_{3}$N ($J=10-9$)}} \\
		n1 & 3.11 (2) & 1.20 (5) & 1.24 (6) & 1.59 (10) & ... \\
		n2 & 4.84 (2) & 2.47 (8) & 0.99 (4) & 2.61 (13) & ...\\
		n3 & 4.69 (1) & 4.08 (8) & 0.79 (2) & 3.42 (10) & ... \\
		n4 & 4.91 (2) & 2.50 (6) & 1.30 (4) & 3.45 (13) & ... \\
		n5 & 6.20 (1) & 2.03 (5) & 1.32 (3) & 2.84 (10) & ... \\
		s1 & 8.22 (1) & 4.80 (4) & 1.87 (2) & 9.55 (12) & ... \\
		s2 & 7.58 (2) & 3.04 (4) & 2.06 (3) & 6.66 (15)& ... \\
		s3 & 7.48 (3) & 1.90 (5) & 2.54 (8) & 5.1 (2) & ... \\
		s4 & 6.58 (4) & 0.86 (3) & 2.61 (10) & 2.40 (12) & ... \\
		s5 & 5.97 (2) & 1.90 (6) & 1.04 (4) & 2.10 (10) & ... \\
		\hline
		\multicolumn{6}{l}{1. Numbers in the parentheses are the standard errors of the Gaussian fit, } \\
		\multicolumn{6}{l}{expressed in units of the last significant digits.}\\
		\multicolumn{6}{l}{2. The rms noise levels for HC$_{3}$N spectra were smaller than 100 mK.}\\
		
	\end{tabular}
\end{table}

\begin{table*}
	\centering
	\caption{Spectral line parameters of N$_{2}$H$^{+}$ ($J=1-0$) at 10 positions}
	\label{tab:gaussN2H}
	\begin{tabular}{ccccccc} 
		\hline
		Position  & Transition & $V_{\rm {LSR}}$ & $T_{\rm {mb}}$  & $\Delta v$ & $\int T_{\rm {mb}} dv$ & rms \\
		    & ($F_{1}'-F_{1}$) & (\,km\,s$^{-1}$) &  (K) & (\,km\,s$^{-1}$)  & (K \,km\,s$^{-1}$) & (mK) \\     
		\hline
		       & $0-1$ & & 1.09 (6) & 1.14 (7) & 1.32 (10) & \\
		 n1  & $2-1$ & 2.40 (2) & 2.95 (4) & 2.10 (3) & 6.59 (14) & 19.3 \\
		       & $1-1$ & & 2.06 (5) & 1.59 (4) & 3.49 (12) & \\
		       & $0-1$ &   & 1.00 (7) & 1.32 (10) & 1.41 (15) & \\
		n2   & $2-1$ & 4.08 (2) & 2.74 (5) & 2.29 (5) & 6.67 (19) & 19.4 \\  
		       & $1-1$ &  &  1.84 (6) & 1.90 (7) & 3.72 (18) & \\
		       & $0-1$ &   & 1.33 (9) & 0.91 (7) & 1.29 (14) & \\
		n3   & $2-1$ & 4.11 (2) & 3.05 (6) & 2.05 (5) & 6.6 (2) & 20.6 \\     
		       & $1-1$ & &  2.21 (7) & 1.46 (6) & 6.43 (17) & \\
		       & $0-1$ &   & 1.03 (3) & 1.70 (6) & 1.86 (8) & \\
		n4   & $2-1$ & 4.32 (1) & 2.71 (3) & 2.54 (3) & 7.32 (10) & 20.7 \\      
		       & $1-1$ & & 1.88 (3) & 2.18 (4) & 4.35 (10) & \\
		       & $0-1$ &   & 2.30 (8) & 1.14 (5) & 2.80 (15) & \\
		n5   & $2-1$ & 5.25 (1) & 5.59 (6) & 2.21 (3) & 13.1 (2) & 20.8 \\
		       & $1-1$ & & 4.21 (7) & 1.69 (3) & 7.60 (18) & \\
		       & $0-1$ &   & 3.73 (8) & 1.79 (5) & 7.1 (3) & \\
		s1   & $2-1$ & 7.56 (1) & 10.93 (7) & 2.63 (2) & 30.5 (3) & 26.6 \\  
		       & $1-1$ & & 7.27 (8) & 2.25 (3) & 17.4 (3) & \\
		       & $0-1$ &   & 3.47 (14) & 1.38 (6) & 5.1(3) & \\
		 s2  & $2-1$ & 6.61 (1) & 10.17 (11) & 2.26 (3) & 24.4 (4) & 20.5 \\     
		       & $1-1$ & & 6.42 (12) & 1.90 (4) & 13.0 (4) & \\
		       & $0-1$ & & 2.13 (14) & 2.6 (2) & 6.0 (6) & \\
		 s3  & $2-1$ & 6.14 (3) & 6.95 (14) & 3.01 (7) & 22.3 (7) & 21.1 \\
		       & $1-1$ & & 4.86 (14) & 2.77 (9) & 14.3 (6) & \\
		        & $0-1$ & & 0.83 (4) & 1.83 (9) & 1.63 (11) & \\ 
		  s4  & $2-1$ &  6.04 (2) & 2.70 (3) & 2.66 (3) & 7.64 (13) & 20.9 \\
		        & $1-1$ & & 1.79 (3) & 2.31 (5) & 4.41 (12) & \\
		        & $0-1$ & & 3.07 (14) & 1.12 (6) & 3.7 (3) & \\
		   s5  & $2-1$ & 5.35 (2) & 5.58 (9) & 2.41 (5) & 14.3 (4) & 21.0 \\   
		         & $1-1$ & & 4.21 (10) & 1.99 (6) & 8.9 (3) & \\   
		\hline
		\multicolumn{7}{l}{Note: Numbers in parentheses are the standard errors of the Gaussian fit, } \\
		\multicolumn{7}{l}{expressed in units of the last significant digits.}\\
	\end{tabular}
\end{table*}

\subsection{Analyses} \label{sec:ana}

We derived the column densities and excitation temperatures of N$_{2}$H$^{+}$ at the 10 positions from a local thermodynamic equilibrium (LTE) analysis, following the procedure and using the formulae derived in Appendix B1 of \citet{2006ApJ...653.1369F}.
The excitation temperatures and column densities are determined simultaneously by a least-square fit using the hyperfine components of N$_{2}$H$^{+}$.
We derived the optical depth as a function of velocity using the formula
\begin{equation}
\tau(v) = \tau_{\rm {tot}} \sum^{7}_{i=1} s_{i} {\rm {exp}} \Biggl[ -4 {\rm {ln}}(2) \biggl(\frac{v - V_{\rm {sys}}}{\Delta v} \biggr)^2 \Biggr],
\end{equation}
where $\tau_{\rm {tot}}$ is the total optical depth, $s_{i}$ is the normalized relative intensity of the $i$th hyperfine component, and $\Delta v$ is the line width (FWHM, Table~\ref{tab:gaussN2H}) of the $i$th component.
We used the $V_{\rm {LSR}}(F_{1}=2-1$) values as $V_{\rm {sys}}$, because this line includes the centroid component.
The column density is derived with the formula
\begin{equation}
N = 3.30 \times 10^{11} \frac{(T_{\rm {ex}}+0.75)}{1-e^{-4.47/T_{\rm {ex}}}} \tau_{\rm {tot}} \Delta v_{int}.
\end{equation}
We assume that all of the hyperfine components have the same thermal and non-thermal line broadenings. 
This assumption can be applicable because all of these lines should come from the same emission region.
We used the line width of $F_{1}=0-1$ as the intrinsic line width ($\Delta v_{int}$), because this component is not blended and consists of only one component.
The derived excitation temperature, total optical depth and column density are summarised in Table~\ref{tab:anaN2H}.

\begin{table}
	\centering
	\caption{Excitation temperature, total optical depth and column density of N$_{2}$H$^{+}$ at 10 positions}
	\label{tab:anaN2H}
	\begin{tabular}{cccc} 
		\hline
		Position  & $T_{\rm ex}$ (K) & $\tau_{\rm {tot}}$ & $N$ ($\times 10^{13}$ cm$^{-2}$) \\
		\hline
		n1 & 12.0 (1.1) & 1.2 (1) & 1.84 (16) \\
		n2 & 11.2 (1.3) & 1.2 (1) &1.9 (2) \\
		n3 & 14.7 (1.8) & 1.1 (1) & 2.0 (2) \\
		n4 &  7.3 (0.4) & 2.5 (1) & 2.47 (13) \\
		n5 & 26.2 (1.6) & 0.95 (6) & 6.2 (4) \\
		s1 & 22.0 (0.8) & 2.0 (1) & 14.7 (6) \\
		s2 & 25.8 (1.8) & 1.5 (1) & 11.4 (8) \\
		s3 & 13.4 (1.6) & 2.1 (2) & 9.1 (1.0) \\
		s4 & 11.4 (0.8) & 0.95 (7) & 2.15 (16) \\
		s5 & 18.6 (1.5) & 2.0 (2) & 6.7 (6) \\
		\hline
		\multicolumn{4}{l}{Note: Numbers in parentheses are the standard errors} \\
		\multicolumn{4}{l}{expressed in units of the last significant digits.}\\
	\end{tabular}
\end{table}

We derived column densities of CCS and HC$_{3}$N assuming the LTE condition, using the formulae \citep{1999ApJ...517..209G}: 
\begin{equation} \label{tau}
\tau = - {\mathrm {ln}} \left[1- \frac{T_{\rm mb} }{\left\{J(T_{\rm {ex}}) - J(T_{\rm {bg}}) \right\}} \right],  
\end{equation}
where
\begin{equation} \label{tem}
J(T) = \frac{h\nu}{k}\Bigl\{\exp\Bigl(\frac{h\nu}{kT}\Bigr) -1\Bigr\} ^{-1},
\end{equation}  
and
\begin{equation} \label{col}
N = \tau \frac{3h\Delta v}{8\pi ^3S}\sqrt{\frac{\pi}{4\mathrm {ln}2}}Q\frac{1}{\mu ^2}\frac{1}{J_{\rm {lower}}+1}\exp\Bigl(\frac{E_{\rm {lower}}}{kT_{\rm {ex}}}\Bigr)\Bigl\{1-\exp\Bigl(-\frac{h\nu }{kT_{\rm {ex}}}\Bigr)\Bigr\} ^{-1}.
\end{equation}  
In Equation (\ref{tau}), $\tau$ and $T_{\rm mb}$ denote the optical depth and peak intensity, respectively, while
$T_{\rm{ex}}$ and $T_{\rm {bg}}$ are the excitation temperature and the cosmic microwave background temperature ($\simeq 2.725$ K), respectively.
We calculated the optical depth and column density assuming that excitation temperatures of 10, 15, 20, 25 K, by taking into consideration the derived excitation temperatures of N$_{2}$H$^{+}$.
$J$($T$) in Equation (\ref{tem}) is the effective temperature equivalent to that in the Rayleigh-Jeans law.
In Equation (\ref{col}), {\it N}, $\Delta v$, $S$, $Q$, $\mu$, and $E_{\rm {lower}}$ denote the column density, line width (FWHM), line strength, partition function, permanent electric dipole moment, and energy of the lower rotational energy level, respectively. 
The rotational constant and permanent electric dipole moment of CCS are 6477.75 MHz and 2.88 Debye, and those of HC$_{3}$N are  4549.059 MHz and 3.73172 Debye\footnote{Splatalogue database; \url{https://www.cv.nrao.edu/php/splat/}}.
We used $T_{\rm mb}$ and the line width obtained from the Gaussian fitting (Table \ref{tab:gaussCCS}).
The derived column densities of CCS and HC$_{3}$N for each position and each assumed excitation temperature are summarised in Table~\ref{tab:anacarbon}.
We also derived the column density ratios of $N$(N$_{2}$H$^{+}$)/$N$(carbon-chain species) at each position as summarised in Table~\ref{tab:anacarbon}.

\begin{landscape}
\begin{table}
	\centering
	\caption{Column density of CCS and HC$_{3}$N and $N$(N$_{2}$H$^{+}$)/$N$(carbon-chain species) ratio at 10 positions}
	\label{tab:anacarbon}
	\begin{tabular}{ccccccccccc} 
		\hline
		Position & $N$(CCS) & $N$(CCS) & $N$(CCS) & $N$(CCS) & $N$(N$_{2}$H$^{+}$)/$N$(CCS)$^{a}$ & $N$(HC$_{3}$N) & $N$(HC$_{3}$N) & $N$(HC$_{3}$N) & $N$(HC$_{3}$N) & $N$(N$_{2}$H$^{+}$)/$N$(HC$_{3}$N)$^{a}$ \\
		  & ($\times 10^{13}$ cm$^{-2}$) & ($\times 10^{13}$ cm$^{-2}$) & ($\times 10^{13}$ cm$^{-2}$) &($\times 10^{13}$ cm$^{-2}$) & & ($\times 10^{13}$ cm$^{-2}$) & ($\times 10^{13}$ cm$^{-2}$) & ($\times 10^{13}$ cm$^{-2}$) & ($\times 10^{13}$ cm$^{-2}$) & \\
		\hline
		n1 & 0.35 (9) & 0.29 (6) & 0.29 (7) & 0.31 (8) & 5.3 (5.3 -- 6.0) & 1.08 (7) & 0.71 (5) & 0.63(4) & 0.62 (4) & 1.7 (1.7 -- 3.0) \\
		n2 & 0.59 (10) & 0.50 (9) & 0.50 (9) &  0.52 (9) & 3.2 (3.2 -- 3.8) & 2.0 (1) & 1.23 (6) & 1.08 (5) & 1.05 (5) & 1.0 (1.0 -- 1.8) \\
		n3 & 0.85 (11) & 0.71 (9) & 0.71 (9) & 0.74 (10) & 2.8 (2.3 --2.8) & 3.19 (9) & 1.77 (5) & 1.50 (4) & 1.44 (4) & 1.1 (0.6 -- 1.4) \\
		n4 & 1.1 (2) & 0.9 (2) & 0.9 (2) & 1.0 (2) & 2.3 (2.3 -- 2.7) & 2.63 (10) & 1.63(6) & 1.43 (5) & 1.39 (5) & 0.9 (0.9 -- 1.8) \\
		n5 & 1.25 (13) & 1.04 (10) & 1.04 (10) & 1.10 (11) & 5.6 (4.9 -- 5.9) & 2.07 (7) & 1.32 (4) & 1.16 (4) & 1.13 (4) & 5.4 (3.0--5.4) \\
		s1 & 2.02 (19) & 1.68 (16) & 1.68 (16) & 1.76 (17) &  8.7 (7.3 -- 8.7) & 10.1 (1) & 5.17 (6) & 4.30 (5) & 4.09 (5) & 3.4 (1.5 -- 3.6) \\
		s2 & 1.01 (13) & 0.84 (11) & 0.84 (11) & 0.88 (11) & 12.9 (11.3 -- 13.5) & 5.41 (12) & 3.25 (7) & 2.81 (6) & 2.72 (6) & 4.2 (2.1 -- 4.2) \\
		s3 & 2.2 (3) & 1.8 (3) & 1.8 (3) & 1.9 (3) & 4.9 (4.1 -- 5.0) & 3.70 (15) & 2.37 (9) & 2.09 (8) & 2.04 (8) & 3.8 (2.5 -- 4.5) \\
		s4 & 0.89 (18) & 0.74 (15) & 0.74 (15) & 0.78 (16) & 2.4 (2.4 -- 2.9) & 1.58 (8) & 1.05 (5) & 0.94 (5) & 0.93 (5) & 1.4 (1.4 -- 2.3) \\
 		s5 & 1.45 (10) & 1.20 (8) & 1.20 (8) & 1.25 (8) & 5.6 (4.6 -- 5.6) & 1.51 (7) & 0.97 (5) & 0.85 (4) & 0.83 (4) & 7.8 (4.4 -- 8.0) \\
\multicolumn{11}{l}{} \\
		Note & $T_{\rm{ex}}=10$ K & $T_{\rm{ex}}=15$ K & $T_{\rm{ex}}=20$ K & $T_{\rm{ex}}=25$ K &  & $T_{\rm{ex}}=10$ K & $T_{\rm{ex}}=15$ K & $T_{\rm{ex}}=20$ K & $T_{\rm{ex}}=25$ K & \\
		\hline
		\multicolumn{11}{l}{Note: Numbers in parentheses are the standard errors expressed in units of the last significant digits.} \\
		\multicolumn{11}{l}{$^{a}$ Values are derived using $N$(carbon-chain species) with the $T_{\rm{ex}}$ closest to $T_{\rm ex}$(N$_{2}$H$^{+}$) for each position. The ranges in parentheses are for the four assumed excitation temperatures.}\\
	\end{tabular}
\end{table}
\end{landscape}

\section{Discussion}

\subsection{The $N$(N$_{2}$H$^{+}$)/$N$(carbon-chain species) Column Density Ratios} \label{sec:d1}

The $N$(N-bearing species)/$N$(carbon-chain species) ratios are classically known as chemical evolutionary indicators in low-mass star-forming regions \citep[e.g.,][]{1992ApJ...392..551S,1998ApJ...506..743B}.
For example, \citet{1998ApJ...506..743B} found that the $N$(N$_{2}$H$^{+}$)/$N$(CCS) column density ratio is lower in starless cores than in star-forming cores by a factor of 2.
The average values are derived to be $\simeq 1.1$ for cores with stars and $\simeq 0.5$ for starless cores, respectively.
In the case of high-mass star-forming regions, \citet{2019ApJ...872..154T} found that the $N$(N$_{2}$H$^{+}$)/$N$(HC$_{3}$N) column density ratio tends to decrease from high-mass starless cores to high-mass protostellar objects.
In this study, we investigate the $N$(N$_{2}$H$^{+}$)/$N$(CCS) and $N$(N$_{2}$H$^{+}$)/$N$(HC$_{3}$N) column density ratios at the 10 positions in order to test whether these ratios can be used as chemical evolutionary indicators. 

Figs.~\ref{fig:carbon}a and ~\ref{fig:carbon}b show plots of $N$(CCS) vs. $N$(N$_{2}$H$^{+}$)/$N$(CCS) and $N$(HC$_{3}$N) vs. $N$(N$_{2}$H$^{+}$)/$N$(HC$_{3}$N), respectively, using points from NGC2264.
Fig.~\ref{fig:carbon}c shows the plot of $N$(N$_{2}$H$^{+}$)/$N$(HC$_{3}$N) vs. $N$(N$_{2}$H$^{+}$)/$N$(CCS) at the 10 positions.
Figs.~\ref{fig:carbon}d and ~\ref{fig:carbon}e show the relationships between the excitation temperature of N$_{2}$H$^{+}$, $T_{\rm {ex}}$(N$_{2}$H$^{+}$), and the $N$(N$_{2}$H$^{+}$)/$N$(CCS) and $N$(N$_{2}$H$^{+}$)/$N$(HC$_{3}$N) ratios, respectively.
The s4 position differs from the others because it has no $850 \mu$m dust continuum clump \citep{2015MNRAS.453.2006B}, while this position seems to be in a molecular outflow \citep{2012MNRAS.423.1127B}.
The large line width of HC$_{3}$N at s4 may support a possibility of the origin of a molecular outflow.
Given its uniqueness, s4 is plotted as an open red circle.

As a general trend, the $N$(N$_{2}$H$^{+}$)/$N$(CCS) values plotted for NGC2264 in Fig.~\ref{fig:carbon}a lie higher than the average value in the low-mass star-forming cores studied by \citet{1998ApJ...506..743B}.
In fact, the average $N$(N$_{2}$H$^{+}$)/$N$(CCS) values for positions in NGC2264 and for the low-mass star-forming cores are 5.4 and 1.1, respectively.
This may imply that the clumps in NGC2264 are more chemically evolved than those previously studied in low-mass star-forming regions. 

We further discuss the characteristics of the $N$(N$_{2}$H$^{+}$)/$N$(CCS) and $N$(N$_{2}$H$^{+}$)/$N$(HC$_{3}$N) ratios in NGC2264 in the following subsections.

\subsubsection{Relationships with the Excitation Temperature of N$_{2}$H$^{+}$} \label{sec:d1-2}

Firstly, we investigate whether the $N$(N$_{2}$H$^{+}$)/$N$(CCS) and $N$(N$_{2}$H$^{+}$)/$N$(HC$_{3}$N) ratios are chemical evolutionary indicators for clumps in the NGC2264 cluster-forming regions. 
In order to do this, we use the excitation temperature of N$_{2}$H$^{+}$, the abscissa in Figs.~\ref{fig:carbon}d and \ref{fig:carbon}e, as a tracer of clump temperature.  
The N$_{2}$H$^{+}$ excitation temperature is expected to increase with clump evolution, in which case it is considered as a physical evolutionary indicator.

A statistical approach was used for determining the correlation between excitation temperature and abundance ratios.
The Kendall's rank correlation method was applied in this paper.
Table~\ref{tab:kendall} summarises the probability ($p$) and the Kendall's rank correlation coefficient ($\tau$).
Because the s2 position seems to be significantly affected by IRS1, we conducted the test for the two cases: (1) all of the 10 positions and (2) 9 positions except for the s2 position.

In the case of (1), the probability that the $N$(N$_{2}$H$^{+}$)/$N$(CCS) ratio and $T_{\rm {ex}}$ are not correlated ($p$-value) is 0.71\%, and the Kendall's rank correlation coefficient ($\tau$) is $+0.67$.
Hence, there is a positive correlation between these parameters.
The corresponding $p$-value and $\tau$ value for the relationship between the $N$(N$_{2}$H$^{+}$)/$N$(HC$_{3}$N) ratio and $T_{\rm {ex}}$ are 0.47\% and $+0.69$, respectively.
Thus, the $N$(N$_{2}$H$^{+}$)/$N$(HC$_{3}$N) ratio also has a positive correlation with $T_{\rm {ex}}$.
If we exclude the s2 position, the corresponding probabilities are slightly higher, but the conclusion that there are positive correlations between the column density ratios and $T_{\rm {ex}}$ does not change.

Since the N$_{2}$H$^{+}$ excitation temperature is considered to increase with clump evolution, the positive correlations between the column density ratios and the N$_{2}$H$^{+}$ excitation temperature mean that the $N$(N$_{2}$H$^{+}$)/$N$(CCS) and $N$(N$_{2}$H$^{+}$)/$N$(HC$_{3}$N) ratios increase as the clump evolves.
Such a characteristic is similar to that found in low-mass star-forming cores \citep{1992ApJ...392..551S,1998ApJ...506..743B}.
We conclude that these column density ratios are chemical evolutionary indicators of clumps in the cluster-forming regions.

\begin{table}
	\centering
	\caption{Summary of the Kendall's rank correlation statics}
	\label{tab:kendall}
	\begin{tabular}{lcc} 
		\hline
		 Pair of parameters & $p$ & $\tau$ \\
		\hline
		\multicolumn{3}{l}{(1) All 10 Positions} \\
		$N$(N$_{2}$H$^{+}$)/$N$(CCS) vs. $T_{\rm {ex}}$ & 0.71\% & $+0.67$ \\
		$N$(N$_{2}$H$^{+}$)/$N$(HC$_{3}$N) vs. $T_{\rm {ex}}$ & 0.47\% & $+0.69$ \\
		\multicolumn{3}{l}{(2) Except for s2} \\
		$N$(N$_{2}$H$^{+}$)/$N$(CCS) vs. $T_{\rm {ex}}$ & 1.59\% & $+0.65$ \\
		$N$(N$_{2}$H$^{+}$)/$N$(HC$_{3}$N) vs. $T_{\rm {ex}}$ & 1.27\% & $+0.67$ \\
	       \hline	
	       \multicolumn{3}{l}{Note: $p$ and $\tau$ represent the probability that there is no} \\
	       \multicolumn{3}{l}{correlation between the two parameters and the} \\
	       \multicolumn{3}{l}{Kendall's rank correlation coefficient, respectively.} \\
	\end{tabular}
\end{table}

\subsubsection{Characteristics in the South Region} \label{sec:d1-1}

We compare the column density ratios among clumps in the South region in this subsection.
The s2 position shows a relatively high $N$(N$_{2}$H$^{+}$)/$N$(CCS) value, whereas the $N$(N$_{2}$H$^{+}$)/$N$(HC$_{3}$N) ratio is not higher than the other positions (Fig.~\ref{fig:carbon}b).
The s1 position also shows  similar features to s2, which can be seen in Fig.~\ref{fig:carbon}c.
These results indicate that CCS is deficient at the s1 and s2 positions, while HC$_{3}$N is relatively abundant.
This suggests that the CCS molecules are efficiently destroyed, while the HC$_{3}$N molecules can survive in these regions.
Because the observed lines of CCS and HC$_{3}$N have similar excitation energies, as listed in Table~\ref{tab:line1}, the difference between CCS and HC$_{3}$N probably does not originate from the excitation conditions.

The s1 and s2 positions seem to be significantly affected by the B2-type star, IRS1.
Around IRS1, the strong UV radiation could efficiently destroy CCS.
On the other hand, HC$_{3}$N is relatively stable and robust against the harsh interstellar environment and thus it can resist the exposure to UV radiation \citep{1995Icar..115..119C}. 
Therefore, if HC$_{3}$N is destroyed to some extent by the effects from IRS1, it is difficult to recognize the effect with the angular resolution of the single-dish telescope. 

\subsubsection{Evolution in the North Region} \label{sec:d1-3}

In this subsection, we discuss evolution in the North region, where the effects from IRS1 seem to be small, using the chemical evolutionary indicators.

\begin{figure*}
       \centering
	\includegraphics[scale=0.8, bb=0 10 477 687]{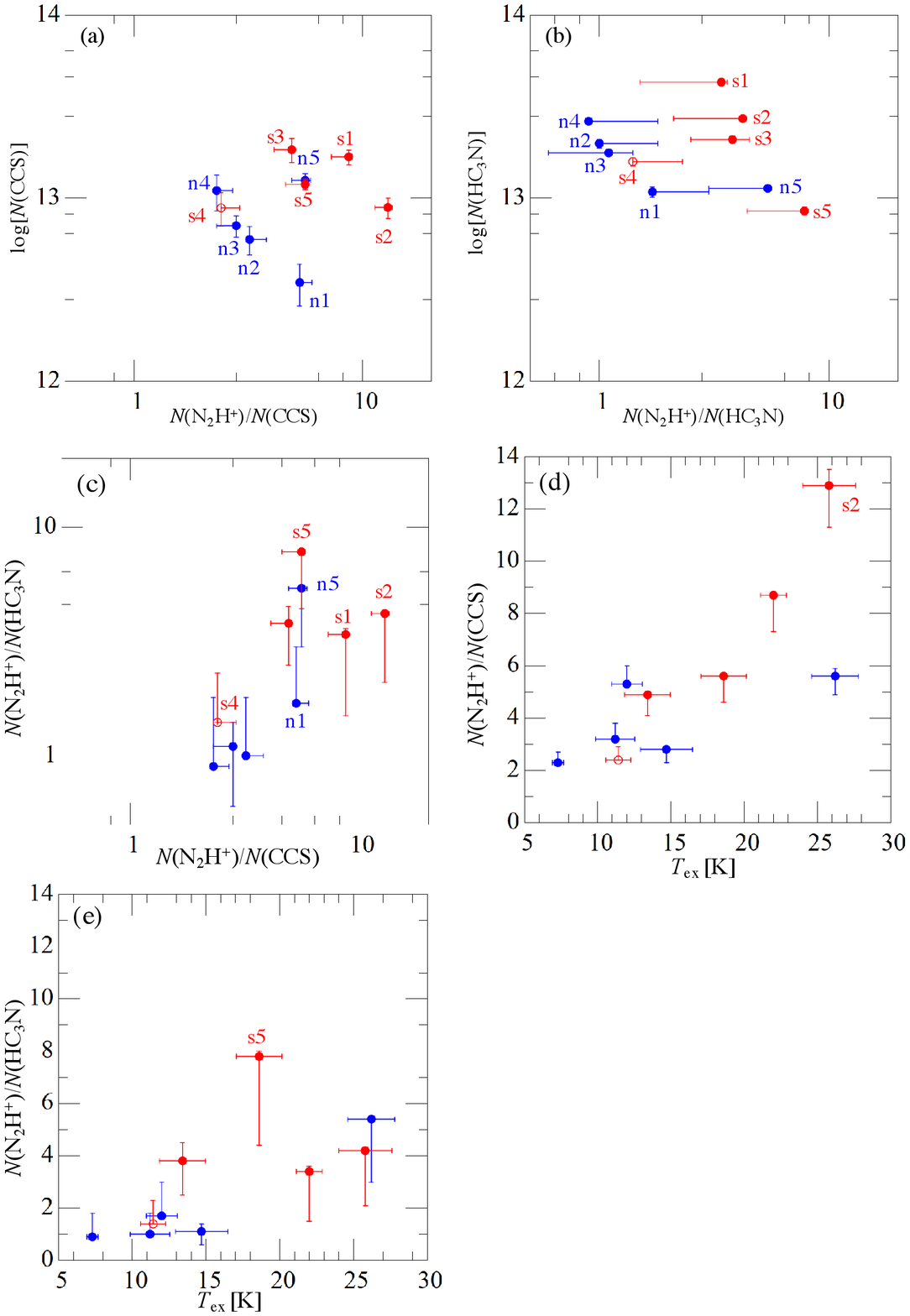}
       \caption{Comparisons of the $N$(N$_{2}$H$^{+}$)/$N$(carbon-chain species) ratios among 10 positions: (a) $N$(CCS) vs. $N$(N$_{2}$H$^{+}$)/$N$(CCS), (b) $N$(HC$_{3}$N) vs. $N$(N$_{2}$H$^{+}$)/$N$(HC$_{3}$N), (c) $N$(N$_{2}$H$^{+}$)/$N$(HC$_{3}$N) vs. $N$(N$_{2}$H$^{+}$)/$N$(CCS), (d) $N$(N$_{2}$H$^{+}$)/$N$(CCS) vs. $T_{\rm {ex}}$(N$_{2}$H$^{+}$), and (f) $N$(N$_{2}$H$^{+}$)/$N$(HC$_{3}$N) vs. $T_{\rm {ex}}$(N$_{2}$H$^{+}$). In panel (b), the error bars for $N$(HC$_{3}$N) are actually included, but plot symbol size is larger than the error bars of $N$(HC$_{3}$N). Hence, the error bars are included in circle symbols.The blue and red symbols represent North and South regions, respectively. The s4 position is indicated by the red open circle, because this position is not associated with  $850 \mu$m dust continuum clump but is associated with a molecular outflow. Labelled data points are referred to in the text especially in Section \ref{sec:d1}.}
    \label{fig:carbon}
\end{figure*}

The $N$(N$_{2}$H$^{+}$)/$N$(CCS) values at n1 and n5 are similar to each other ($\approx 5-6$) and they are higher than the north-west positions, n2 -- n4 (Fig.~\ref{fig:carbon}a).
These ratios mean that the n1 and n5 positions are chemically more evolved than the others, from the discussion in Section \ref{sec:d1-2}.

There are two possible interpretations for this tendency.
One is the chemical trend along the filamentary structure.
The ratio seems to decrease from south-east to north-west.
If IRS1 has affected its surrounding and induced star formation, the chemistry may proceed faster around IRS1 and such a chemical differentiation between the south-east and north-west positions in the North region may occur.

The second interpretation of the tendency involves the chemical evolution of the clumps.
We now discuss combining information of protostars.
Both the n1 and n5 positions contain the $850 \mu$m dust continuum clump associated with both Class 0/I and II protostars \citep{2015MNRAS.453.2006B}.
The n2 position is associated with a Class II protostar, whereas the n3 and n4 positions contain Class 0/I protostars \citep{2015MNRAS.453.2006B}.
The clumps with both Class 0/I and II sources show the highest $N$(N$_{2}$H$^{+}$)/$N$(CCS) value, while the clumps associated with Class 0/I protostars show low values, and the clump associated with only a Class II protostar shows an intermediate value.
This trend is independent of the number of the identified YSOs (see Table~\ref{tab:position}).
Since the n1 and n5 positions are considered to be chemically more evolved, we can infer that clumps associated with both Class 0/I and II protostars are more evolved than that associated with only a Class II protostar.
This scenario suggests that a second round of star formation has been occurring in the clumps associated with both Class 0/I and II protostars.

As seen in Fig.~\ref{fig:carbon}b, the $N$(N$_{2}$H$^{+}$)/$N$(HC$_{3}$N) ratio at the n5 position is higher than the other positions.
The values at the three north-west positions (n2 -- n4) are similar to each other, while the value at position n1 is slightly higher than these three positions and marginally consistent with that at n5.
The tendency of this ratio is largely similar to that of the $N$(N$_{2}$H$^{+}$)/$N$(CCS) ratio, as mentioned in the preceding paragraph.
Hence, the $N$(N$_{2}$H$^{+}$)/$N$(HC$_{3}$N) ratio also supports the idea of clump evolution: clumps associated with both Class 0/I and Class II protostars seem to be more evolved than that associated with only a Class II source.

In summary, the $N$(N$_{2}$H$^{+}$)/$N$(CCS) and $N$(N$_{2}$H$^{+}$)/$N$(HC$_{3}$N) ratios seem to reflect both the evolution along the filament and evolution of each clump.

\subsection{The $I$(HC$_{3}$N)/$I$(CH$_{3}$OH) Integrated Intensity Ratio} \label{sec:d2}

Chemical differentiation between COMs-rich and carbon-chain-rich regions has been found around both low- and high-mass protostars, as mentioned in Section \ref{sec:intro}.
Such chemical differentiation seems to suggest a variety of formation processes \citep{2013ChRv..113.8981S, 2019ApJ...881..57T} and environments \citep{2016A&A...592L..11S}.
In this subsection, we investigate such a chemical differentiation using the $I$(HC$_{3}$N)/$I$(CH$_{3}$OH) ratio.

\begin{figure}
       \centering
	\includegraphics[width=6cm]{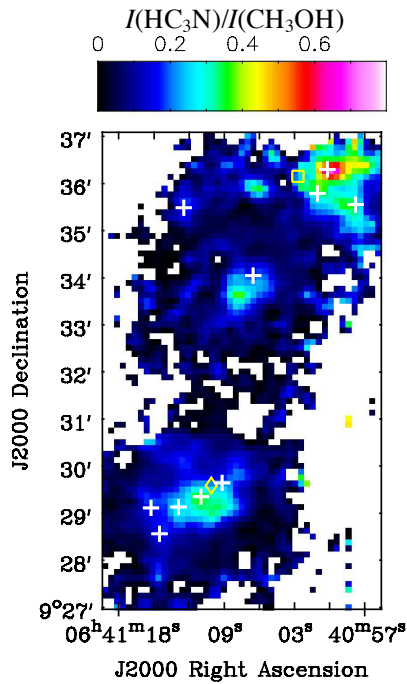}
    \caption{Ratio map of the integrated intensities of HC$_{3}$N and CH$_{3}$OH, $I$(HC$_{3}$N)/$I$(CH$_{3}$OH). The yellow diamond and square indicate the IRS1 and IRS2 positions, respectively. The white cross marks indicate n1$-$n5 and s1$-$s5 positions.}
    \label{fig:ratio}
\end{figure}

We constructed a ratio map of the integrated intensities of the HC$_{3}$N and CH$_{3}$OH emission lines, $I$(HC$_{3}$N)/$I$(CH$_{3}$OH), which is shown in Fig.~\ref{fig:ratio}.
Because the high-excitation-energy line of CH$_{3}$OH has been detected only around IRS1, we only used its low-excitation-energy lines, as shown in Fig.~\ref{fig:mom0}b.  
Table~\ref{tab:ratio} summarises the $I$(HC$_{3}$N)/$I$(CH$_{3}$OH) values at the 10 positions.
The highest $I$(HC$_{3}$N)/$I$(CH$_{3}$OH) ratio is derived to be $\approx 0.7$, and is found at the n3 position.
This value is larger than the other positions by a factor of more than two.
In the South region, the highest value is 0.35 at the s1 position, corresponding to the center of the HC$_{3}$N distribution (Fig.~\ref{fig:mom0}e).

In addition to possessing the highest ratio, the n3 position shows a relatively higher excitation temperature of N$_{2}$H$^{+}$ compared with the n2 and n4 positions, and four YSOs were associated with n3 within the beam size of $18 \arcsec$ \citep{2003yCat.2246....0C}.
Therefore, n3 is not a starless clump and the highest $I$(HC$_{3}$N)/$I$(CH$_{3}$OH) ratio could not be explained by the different chemical evolutionary stages between the n3 and other positions.
The highest $I$(HC$_{3}$N)/$I$(CH$_{3}$OH) ratio toward n3 may imply chemical diversity; the n3 position is a cyanopolyyne-rich and CH$_{3}$OH-poor clump with protostars \citep{2017ApJ...844...68T,2018ApJ...866..150T}.

As mentioned previously, three possible origins of such cyanopolyyne-rich and CH$_{3}$OH-poor conditions have been proposed: a short timescale for prestellar collapse \citep{2013ChRv..113.8981S}, the penetration of the interstellar radiation field \citep{2016A&A...592L..11S}, and a long warm-up stage \citep{2019ApJ...881..57T}.
The n3 position is located at the edge of the filamentary structure (Fig.~\ref{fig:mom0}).
Hence, the effects of the interstellar radiation field may be partially related to the high $I$(HC$_{3}$N)/$I$(CH$_{3}$OH) ratio at the n3 position.
However, both NGC2264-C and NGC2264-D are located in the denser part of the giant molecular cloud \citep{2012MNRAS.423.1127B,2014ApJ...794..124R}. 
The effect of the interstellar radiation field in these cluster regions is still unclear.

Around massive protostars, the effects of  UV radiation from the massive protostars should be more significant than the interstellar radiation field.
If the UV radiation from the massive protostar effects such a chemical differentiation, we would expect the carbon-chain species to be abundant near the massive protostar, because the UV radiation would produce C and/or C$^{+}$, which are precursors of carbon-chain species.
Thus, the UV radiation from the massive protostar is unlikely to be related to the observed $I$(HC$_{3}$N)/$I$(CH$_{3}$OH) ratio.

From our observations, we can evaluate the timescales of prestellar collapse and warm-up stages from the large scale point of view, since the beam size of the Nobeyama 45-m telescope cannot resolve each star-forming core.
The age of NGC2264 is estimated to be $\sim3$ Myr \citep{2004AJ....128.1684S}.
Hence, there was enough time for carbon ions (C$^{+}$) to be converted into CO molecules \citep{2019ApJ...884...167T} before protostars were born in these clumps. 
In that case, CH$_{3}$OH would be formed efficiently by hydrogenation of CO molecules on dust surfaces leading to a CH$_{3}$OH-rich condition, and the observed high $I$(HC$_{3}$N)/$I$(CH$_{3}$OH) ratio at the n3 position could not be explained by the timescale of prestellar collapse.

As mentioned in Section \ref{sec:intro}, the timescale of warm-up stage is considered to be related to the infall velocity \citep{2019ApJ...881..57T}. 
\citet{2006A&A...445..979P} derived the mass infall rates toward the central protostar object in NGC2264-C (South region) and D-MM1 source, which corresponds to the n5 position in this paper, in NGC2264-D (North region) to be $\sim3 \times 10^{-3}$ M$_{\sun}$ yr$^{-1}$ and $\sim1.1 \times 10^{-4}$ M$_{\sun}$ yr$^{-1}$, respectively.
Therefore, the infall velocity is smaller in the North region by nearly one order of magnitude.
It will take longer time to gather gas and dust during the formation of protostars in the North region, leading to the long warm-up period.
In that case, a cyanopolyyne-rich and CH$_{3}$OH-poor condition can be realized \citep{2019ApJ...881..57T}.

In summary, the interstellar radiation field and infall velocity (related to the timescale of the warm-up stage) may also be related to the high $I$(HC$_{3}$N)/$I$(CH$_{3}$OH) intensity ratio at the n3 position with the data obtained by the single-dish telescope.

\begin{table}
	\centering
	\caption{Summary of the $I$(HC$_{3}$N)/$I$(CH$_{3}$OH) ratio}
	\label{tab:ratio}
	\begin{tabular}{cc} 
		\hline
		     position & Value \\
		\hline
		n1 & 0.16 \\
		n2 & 0.32 \\
		n3 & 0.71 \\
		n4 & 0.29 \\
		n5 & 0.19 \\
		s1 & 0.35 \\
		s2 & 0.22 \\
		s3 & 0.26 \\
		s4 & 0.16 \\
		s5 & 0.16 \\
		\hline	
	\end{tabular}
\end{table}		

\section{Conclusions} \label{sec:con}

We have carried out mapping observations of HC$_{3}$N and CH$_{3}$OH toward NGC2264-C and NGC2264-D with the Nobeyama 45-m radio telescope in order to investigate the chemical differentiation between these cluster-forming clumps.
We derived the $I$(HC$_{3}$N)/$I$(CH$_{3}$OH) map, and found the ratio at the north-west position in NGC2264-D to be higher than the other positions by a factor of 2. 
The result may suggest a chemical diversity.
The interstellar radiation field and the small infall velocity, corresponding to a long timescale of warm-up stage,  may be related to the high value at the northern edge of NGC2264-D.

We also observed the molecular emission lines of CCS and N$_{2}$H$^{+}$ toward 10 positions in the two cluster regions,
and derived the $N$(N$_{2}$H$^{+}$)/$N$(CCS) and $N$(N$_{2}$H$^{+}$)/$N$(HC$_{3}$N) column density ratios at these positions.
The $N$(N$_{2}$H$^{+}$)/$N$(CCS) value toward IRS1, the B2-type massive protostar, is much higher than toward the other positions, whereas the $N$(N$_{2}$H$^{+}$)/$N$(HC$_{3}$N) value is comparable with those at the other positions. 
These results may imply that CCS is efficiently destroyed around the massive protostar. 
Furthermore, these ratios have positive correlations with the excitation temperature of N$_{2}$H$^{+}$, which is suggestive of a chemical evolution of clumps.
These column density ratios likely reflect a combination of  evolution occurring both along the filament and in each clump.

\section*{Acknowledgements}

We are deeply grateful to the staff of the Nobeyama Radio Observatory.
The Nobeyama Radio Observatory is a branch of the National Astronomical Observatory of Japan (NAOJ), National Institutes of Natural Science (NINS).
K. T. would like to thank the University of Virginia for providing the funds for her postdoctoral fellowship in the Virginia Initiative on Cosmic Origins (VICO) research program.  
E. H. thanks the National Science Foundation for support of his astrochemistry program through grant AST 1906489. 
Part of this work was supported by NAOJ Research Coordination Committee, NINS, Grant Number 1901-0402.
The National Radio Astronomy Observatory is a facility of the National Science Foundation operated under cooperative agreement by Associated Universities, Inc.





\begin{thebibliography}{99}
\bibitem[\protect\citeauthoryear{Adams}{2010}]{2010ARA&A..48...47A} Adams F.~C., 2010, ARA\&A, 48, 47
\bibitem[\protect\citeauthoryear{Arce et al.}{2007}]{2007prpl.conf..245A} Arce H.~G., Shepherd D., Gueth F., Lee C.-F., Bachiller R., Rosen A., Beuther H., 2007, prpl.conf,  245, prpl.conf
\bibitem[\protect\citeauthoryear{Benson, Caselli \& Myers}{1998}]{1998ApJ...506..743B} Benson P.~J., Caselli P., Myers P.~C., 1998, ApJ, 506, 743
\bibitem[\protect\citeauthoryear{Buckle \& Richer}{2015}]{2015MNRAS.453.2006B} Buckle J.~V., Richer J.~S., 2015, MNRAS, 453, 2006 
\bibitem[\protect\citeauthoryear{Buckle, Richer \& Davis}{2012}]{2012MNRAS.423.1127B} Buckle J.~V., Richer J.~S., Davis C.~J., 2012, MNRAS, 423, 1127 
\bibitem[\protect\citeauthoryear{Caselli \& Ceccarelli}{2012}]{2012A&ARv..20...56C} Caselli P., Ceccarelli C., 2012, A\&ARv, 20, 56
\bibitem[\protect\citeauthoryear{Clarke \& Ferris}{1995}]{1995Icar..115..119C} Clarke D.~W., Ferris J.~P., 1995, Icarus, 115, 119
\bibitem[\protect\citeauthoryear{Cutri et al.}{2003}]{2003yCat.2246....0C} Cutri R.~M., et al., 2003, yCat, II/246
\bibitem[\protect\citeauthoryear{Dahm}{2008}]{2008hsf1.book..966D} Dahm S.~E., 2008, Handbook of Star Forming Regions, Volume I,  966, hsf1.book
\bibitem[\protect\citeauthoryear{Dobashi}{2011}]{2011PASJ...63S...1D} Dobashi K., 2011, PASJ, 63, S1
\bibitem[\protect\citeauthoryear{Furuya, Kitamura \& Shinnaga}{2006}]{2006ApJ...653.1369F} Furuya R.~S., Kitamura Y., Shinnaga H., 2006, ApJ, 653, 1369
\bibitem[\protect\citeauthoryear{Garrod \& Herbst}{2006}]{2006A&A...457..927G} Garrod R.~T., Herbst E., 2006, A\&A, 457, 927
\bibitem[\protect\citeauthoryear{Goldsmith \& Langer}{1999}]{1999ApJ...517..209G} Goldsmith P.~F., Langer W.~D., 1999, ApJ, 517, 209
\bibitem[\protect\citeauthoryear{Hassel, Herbst \& Garrod}{2008}]{2008ApJ...681.1385H} Hassel G.~E., Herbst E., Garrod R.~T., 2008, ApJ, 681, 1385
\bibitem[\protect\citeauthoryear{Herbst \& van Dishoeck}{2009}]{2009ARA&A..47..427H} Herbst E., van Dishoeck E.~F., 2009, ARA\&A, 47, 427
\bibitem[\protect\citeauthoryear{Lada \& Lada}{2003}]{2003ARA&A..41...57L} Lada C.~J., Lada E.~A., 2003, ARA\&A, 41, 57
\bibitem[\protect\citeauthoryear{Minamidani et al.}{2016}]{2016SPIE.9914E..1ZM} Minamidani T., et al., 2016, Millimeter, Submillimeter, and Far-Infrared Detectors and Instrumentation for Astronomy VIII,  99141Z, SPIE.9914
\bibitem[\protect\citeauthoryear{M{\"u}ller et al.}{2005}]{2005JMoSt.742..215M} M{\"u}ller H.~S.~P., Schl{\"o}der F., Stutzki J., Winnewisser G., 2005, JMoSt, 742, 215
\bibitem[\protect\citeauthoryear{Peretto, Andr{\'e} \& Belloche}{2006}]{2006A&A...445..979P} Peretto N., Andr{\'e} P., Belloche A., 2006, A\&A, 445, 979
\bibitem[\protect\citeauthoryear{Rapson et al.}{2014}]{2014ApJ...794..124R} Rapson V.~A., Pipher J.~L., Gutermuth R.~A., Megeath S.~T., Allen T.~S., Myers P.~C., Allen L.~E., 2014, ApJ, 794, 124
\bibitem[\protect\citeauthoryear{Sakai \& Yamamoto}{2013}]{2013ChRv..113.8981S} Sakai N., Yamamoto S., 2013, ChRv, 113, 8981
\bibitem[\protect\citeauthoryear{Sawada et al.}{2008}]{2008PASJ...60..445S} Sawada T., et al., 2008, PASJ, 60, 445
\bibitem[\protect\citeauthoryear{Shimoikura et al.}{2013}]{2013ApJ...768...72S} Shimoikura T., et al., 2013, ApJ, 768, 72
\bibitem[\protect\citeauthoryear{Shimoikura et al.}{2018}]{2018ApJ...855...45S} Shimoikura T., Dobashi K., Nakamura F., Matsumoto T., Hirota T., 2018, ApJ, 855, 45
\bibitem[\protect\citeauthoryear{Skouteris et al.}{2019}]{2019MNRAS.482.3567S} Skouteris D., Balucani N., Ceccarelli C., Faginas Lago N., Codella C., Falcinelli S., Rosi M., 2019, MNRAS, 482, 3567
\bibitem[\protect\citeauthoryear{Spezzano et al.}{2016}]{2016A&A...592L..11S} Spezzano S., Bizzocchi L., Caselli P., Harju J., Br{\"u}nken S., 2016, A\&A, 592, L11
\bibitem[\protect\citeauthoryear{Sung, Bessell \& Chun}{2004}]{2004AJ....128.1684S} Sung H., Bessell M.~S., Chun M.-Y., 2004, AJ, 128, 1684
\bibitem[\protect\citeauthoryear{Suzuki et al.}{1992}]{1992ApJ...392..551S} Suzuki H., Yamamoto S., Ohishi M., Kaifu N., Ishikawa S.-I., Hirahara Y., Takano S., 1992, ApJ, 392, 551
\bibitem[\protect\citeauthoryear{Taniguchi et al.}{2017}]{2017ApJ...844...68T} Taniguchi K., et al., 2017, ApJ, 844, 68
\bibitem[\protect\citeauthoryear{Taniguchi et al.}{2018a}]{2018ApJ...866..150T} Taniguchi K., et al., 2018a, ApJ, 866, 150
\bibitem[\protect\citeauthoryear{Taniguchi et al.}{2019a}]{2019ApJ...881..57T} Taniguchi K., Herbst E., Caselli P., Paulive A., Maffucci D.~M., Saito M., 2019a, ApJ, 881, 57
\bibitem[\protect\citeauthoryear{Taniguchi et al.}{2019b}]{2019ApJ...884...167T} Taniguchi K., Herbst E., Ozeki H., Saito M., 2019b, ApJ, 884, 167
\bibitem[\protect\citeauthoryear{Taniguchi et al.}{2018b}]{2018ApJ...854..133T} Taniguchi K., Saito M., Sridharan T.~K., Minamidani T., 2018b, ApJ, 854, 133
\bibitem[\protect\citeauthoryear{Taniguchi et al.}{2019c}]{2019ApJ...872..154T} Taniguchi K., Saito M., Sridharan T.~K., Minamidani T., 2019c, ApJ, 872, 154
\bibitem[\protect\citeauthoryear{van Dishoeck}{2018}]{2018IAUS..332....3V} van Dishoeck, E.~F.\ 2018, IAU Symposium, 332, 3
\bibitem[\protect\citeauthoryear{V{\'a}zquez-Semadeni, Gonz{\'a}lez-Samaniego \& Col{\'\i}n}{2017}]{2017MNRAS.467.1313V} V{\'a}zquez-Semadeni E., Gonz{\'a}lez-Samaniego A., Col{\'\i}n P., 2017, MNRAS, 467, 1313
\bibitem[\protect\citeauthoryear{Watanabe et al.}{2015}]{2015ApJ...809..162W} Watanabe Y., et al., 2015, ApJ, 809, 162
\bibitem[\protect\citeauthoryear{Watanabe et al.}{2017}]{2017ApJ...847..108W} Watanabe Y., et al., 2017, ApJ, 847, 108
\bibitem[\protect\citeauthoryear{Williams \& Garland}{2002}]{2002ApJ...568..259W} Williams J.~P., Garland C.~A., 2002, ApJ, 568, 259
\end{thebibliography}



\appendix

\section{Spectra of CH$_{3}$OH and HC$_{3}$N at the 10 positions} \label{sec:appen}

\begin{figure*}
	\includegraphics[width=12cm]{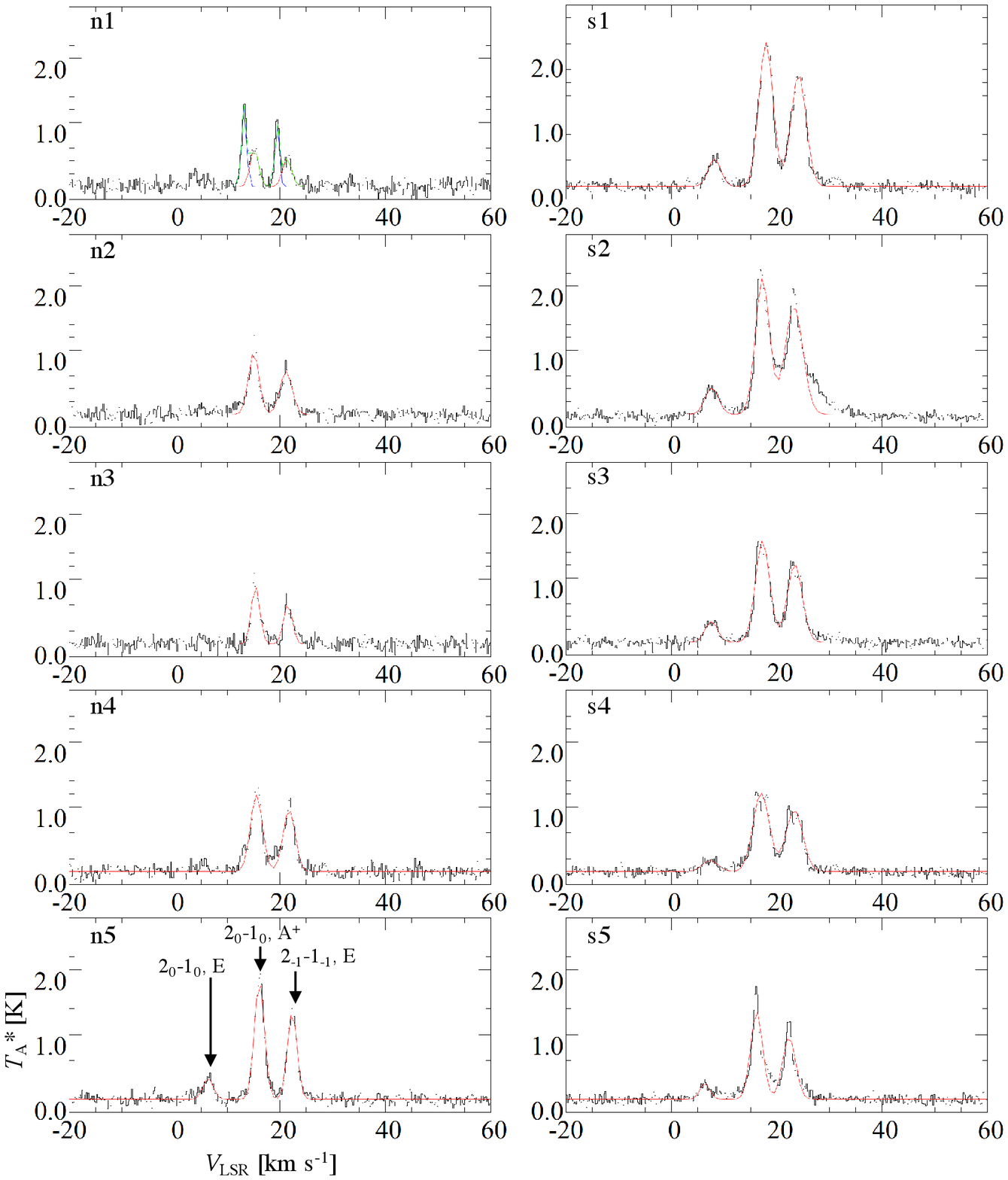}
    \caption{Spectra of CH$_{3}$OH ($2_{0}-1_{0}$, E) toward n1 -- n5 and s1 -- s5 obtained from the OTF data with the beam size of 18\arcsec. 
    In the case of n1, we applied two-velocity-component Gaussian fitting indicated by blue and red curves, and the synthesized spectra of the two components are indicated by green curve.
    The red curves indicate the results of the Gaussian fitting. We also observe the $2_{0}-1_{0}$, A$^{+}$ (96.741371 GHz; $E_{\rm {u}}/k = 6.97$ K) and $2_{-1}-1_{-1}$, E (96.739358 GHz; $E_{\rm {u}}/k = 12.54$ K) lines simultaneously.}
    \label{fig:CH3OHlow}
\end{figure*}

\begin{figure*}
	\includegraphics[width=12cm]{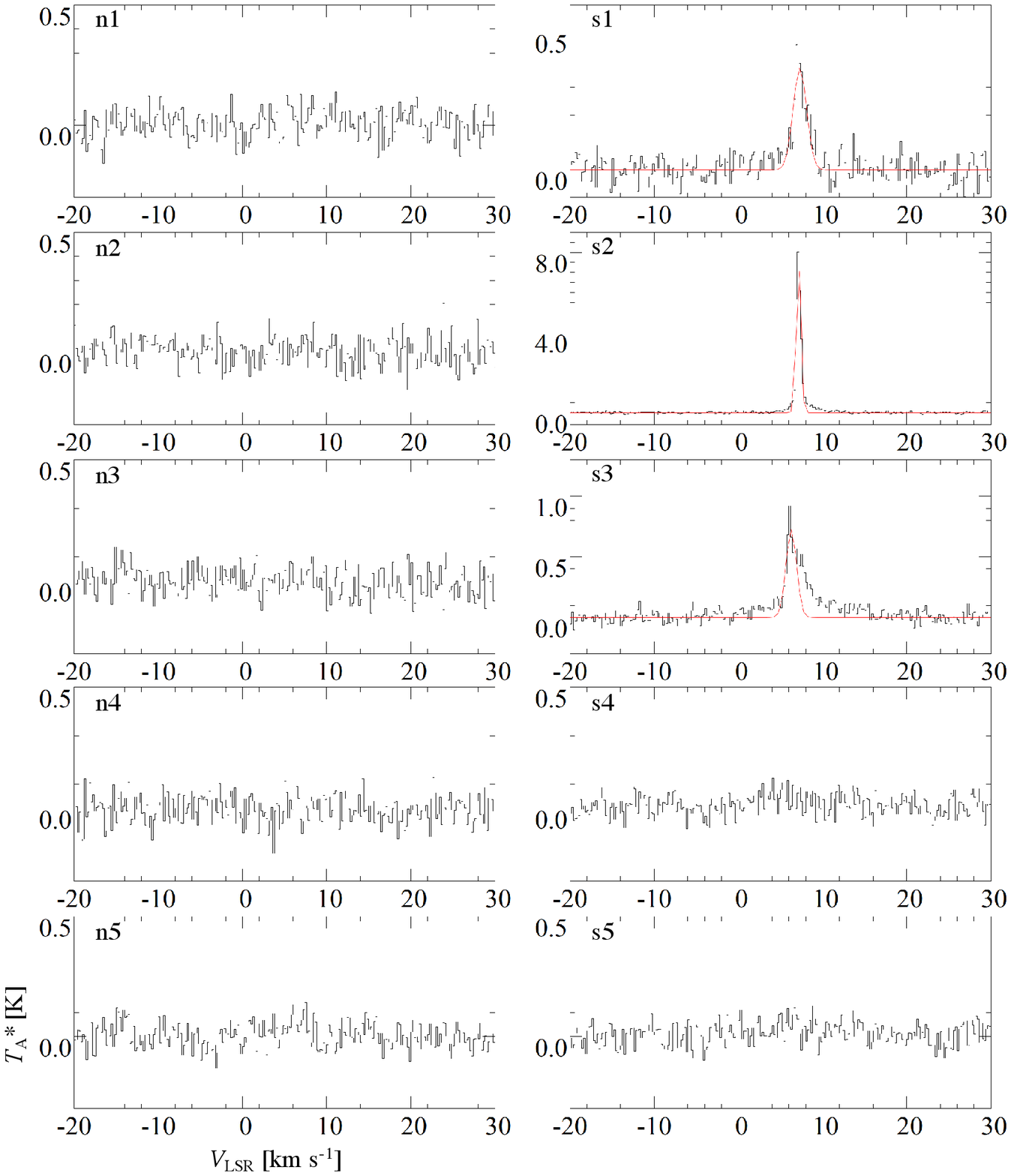}
    \caption{Spectra of CH$_{3}$OH ($8_{0}-7_{1}$, A$^{+}$ ) toward n1 -- n5 and s1 -- s5 obtained from the OTF data with the beam size of 18\arcsec. 
    The red curves indicate the results of the Gaussian fitting. }
    \label{fig:CH3OHhigh}
\end{figure*}

\begin{figure*}
	\includegraphics[width=12cm]{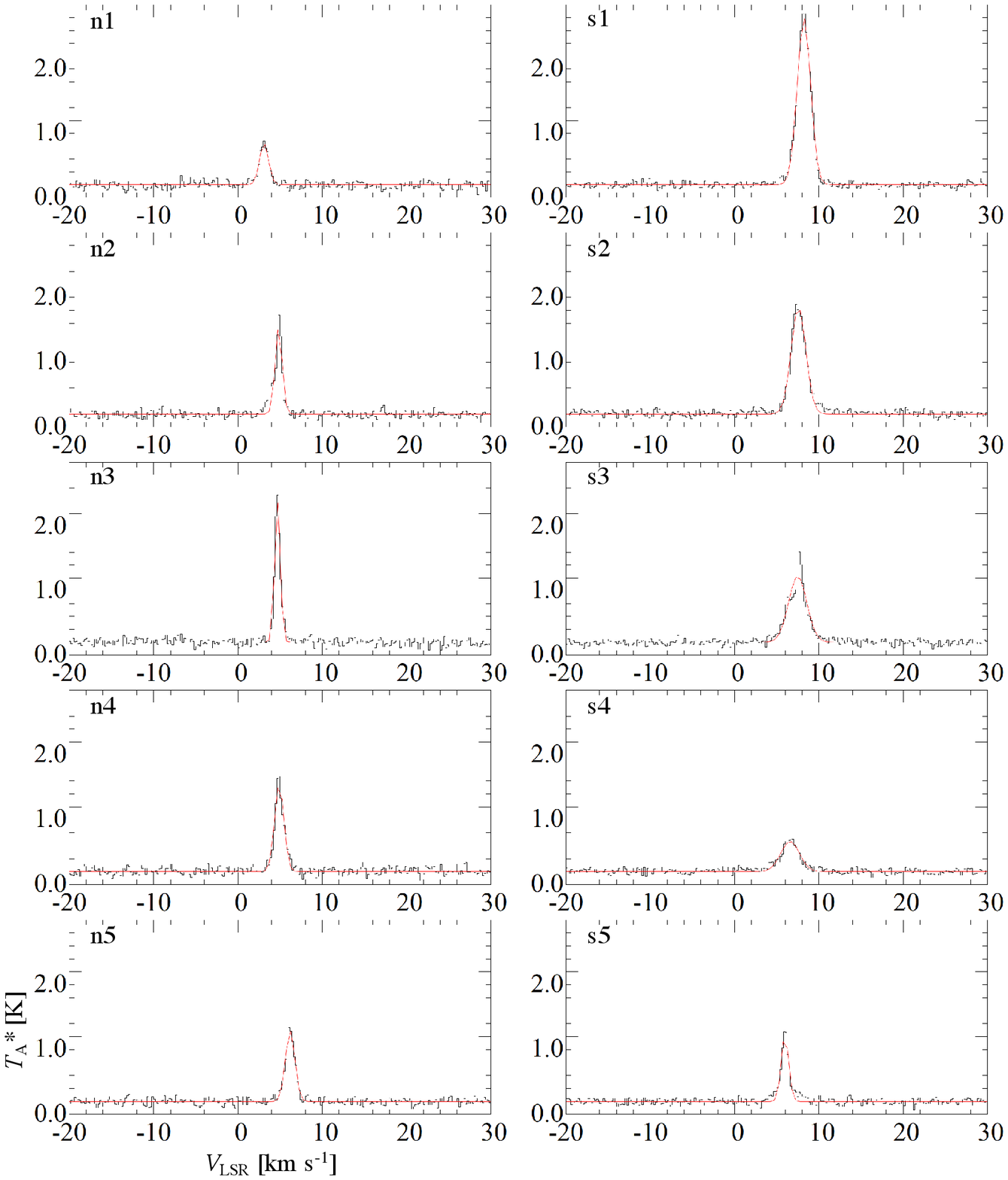}
    \caption{Spectra of HC$_{3}$N ($J=10-9$) toward n1 -- n5 and s1 -- s5 obtained from the OTF data with the beam size of 18\arcsec. 
    The red curves indicate the results of the Gaussian fitting. }
    \label{fig:HC3N}
\end{figure*}

In Figs.\ref{fig:CH3OHlow}, \ref{fig:CH3OHhigh}, and \ref{fig:HC3N}, we show spectra of CH$_{3}$OH (both low- and high-excitation-energy lines) and HC$_{3}$N at the 10 positions obtained from the OTF observations.

\bsp	
\label{lastpage}
\end{document}